\documentclass[twocolumn,secnumarabic,amssymb, nobibnotes, aps, pra]{revtex4-2}

\usepackage{amsmath}
\usepackage{xcolor}
\usepackage{bbold}
\usepackage{ulem}
\usepackage{hyperref}
\usepackage{comment}
\usepackage{natbib}
\usepackage{float}
\usepackage{physics}
\usepackage{graphicx}
\usepackage{algorithm}
\usepackage{algpseudocode}
\hypersetup{
    colorlinks=true,
    linkcolor=red,
    citecolor=magenta,
    filecolor=magenta,
    urlcolor=cyan,
    runcolor=cyan
}

\begin{document}

\title{Statistical analysis of Multipath Entanglement Purification in Quantum Networks}
\author{Rajni Bala$^*$, Md Sohel Mondal$^*$, Siddhartha Santra\\\textit{Department of Physics and Center of excellence in Quantum information, computation, science and technology, Indian Institute of Technology Bombay, Mumbai, Maharashtra 400076, India}\\ $^*$ Equal contribution.}

\begin{abstract}
In quantum networks, a set of entangled states distributed over multiple, alternative, distinct paths between a pair of source-destination nodes can be purified to obtain a higher fidelity entangled state between the nodes. This multipath entanglement purification (MP-EP) strategy can exploit the network's complex structure to strengthen the entanglement connection between node pairs separated by appropriate graph distances. We investigate the network scenarios in which MP-EP outperforms entanglement distribution over single network paths utilising a statistical model of a quantum network and find that MP-EP can be an effective entanglement distribution strategy over a range of node separations determined by the average edge fidelities and probabilities of the network. We find that MP-EP can boost the entanglement connection between suitably separated node pairs to reach fidelities sufficient for a given quantum task thereby increasing the functionality of a quantum network. Further, we provide statistical criteria in terms of network parameters that can determine the regions of the network where MP-EP can be a useful entanglement distribution strategy.
\end{abstract}
\maketitle
\section{Introduction}
Large-scale classical communication networks exhibit complex-network properties that can be harnessed to robustly communicate between arbitrary pairs of source-destination (S-D) nodes. An important complex-network property of large-scale random classical networks is the availability of multiple distinct paths (no common edges) between arbitrary pairs of S-D nodes \cite{boccaletti2006complex,tanenbaum2003computer}. These multiple distinct paths between a given node pair in a classical network can be alternatively used for sending packet-switched data leading to resilient network connectivity even when faced with significant levels of node or link failures \cite{van2010graph,chiesa2021survey}. This observation leads to a natural question: how can we use the similar availability of multiple alternative distinct (MAD) paths in quantum networks for higher fidelity entanglement distribution compared to the use of a single path between a given S-D pair of network nodes?

Quantum networks at sufficiently large scales may be expected to possess the MAD path property between arbitrary node-pairs \cite{wehner2018quantum,leone2021qunet,brekenfeld2020quantum,kimble2008quantum,pirandola2016physics}, akin to their classical counterparts, for two main reasons: First, quantum networks are likely to utilise existing classical communication networks and the internet as the underlying classical control and communication backbone \cite{kozlowski2019towards}- thereby inheriting its topology and associated complex-network properties \cite{wehner2018quantum,Barabasi@network}. Second, the MAD paths are a generic feature \footnote{This fails for pathological topologies such as trees.} of random network topologies which should hold for quantum networks even with topologies that are distinct from those of classical networks\cite{Barabasi@network,kozlowski2019towards}.

In a quantum network, one way to exploit the MAD path feature for  entanglement-based quantum tasks \cite{pant2019routing,kimble2008quantum,wehner2018quantum}, in general, is by establishing end-to-end entanglement, via a given entanglement distribution protocol, separately along each distinct path and then utilising each such path independently for the desired quantum task. In this view, the long-ranged entangled state that can be established along each distinct path provides an alternative quantum channel for providing distributed entanglement as a network resource. For example, to perform quantum key distribution (QKD) between a pair of S-D nodes, the entangled states obtained along distinct paths can be separately and simultaneously used to generate keys with the total key rate being the sum of the rates along all the available paths.

\begin{figure}
    \centering
     \includegraphics[width=\linewidth]{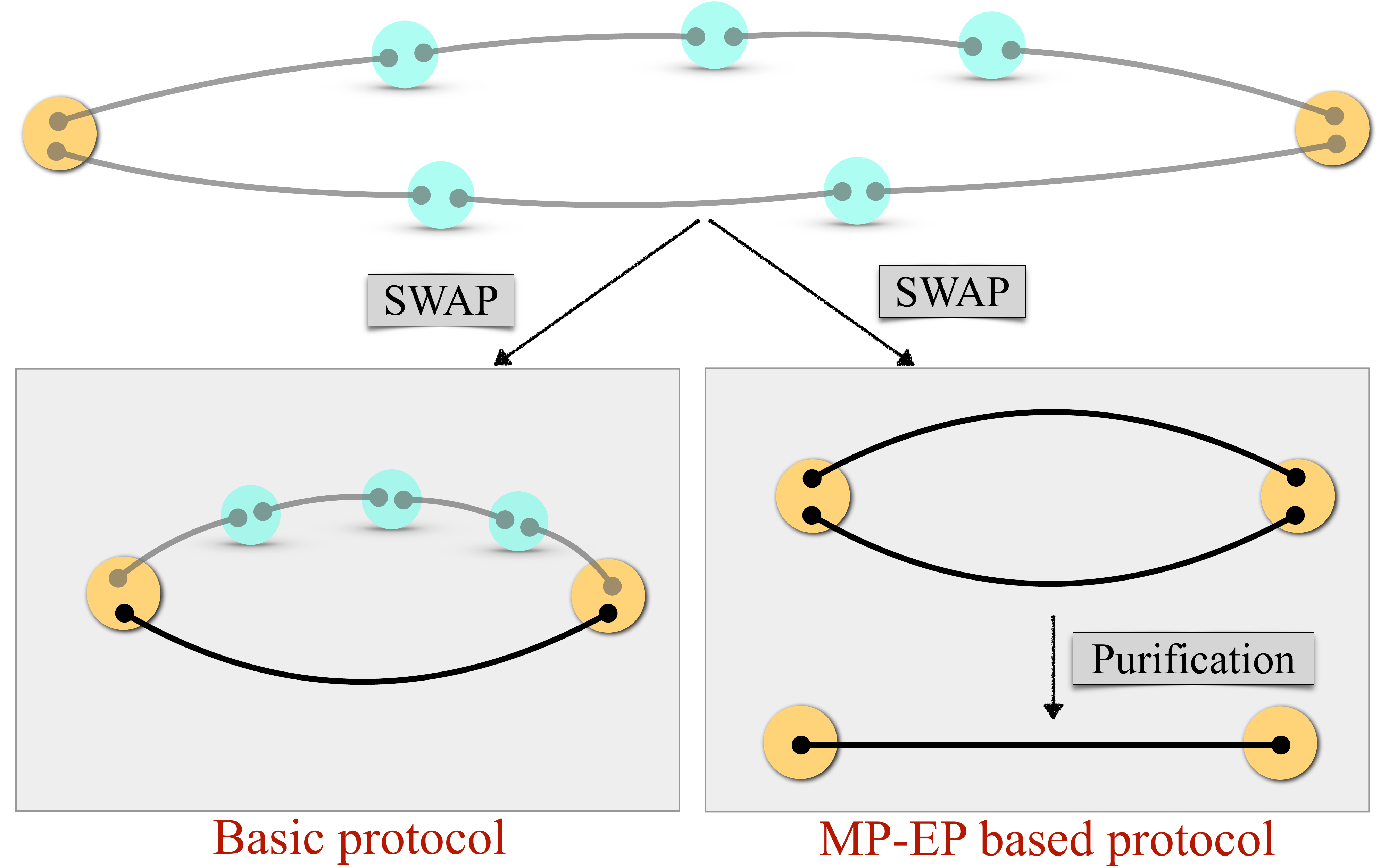}
    \caption{Schemes for single path based (Basic) and multipath based (MP-EP) entanglement distribution protocols. In the basic protocol, shown to the left, 
    entanglement swapping at intermediate nodes along a single path (shortest graph path) between a pair of source-destination nodes is used for entanglement distribution. In the MP-EP protocol, shown to the right, the entangled states established between the the same pair of nodes along two distinct paths (the shortest graph path and an alternate path) are purified to obtain a high fidelity entangled pair. The MP-EP protocol yields a higher effective fidelity of the entanglement connection if the source-destination nodes are appropriately spaced and the two paths have comparable lengths.}
    \label{fig:MEP}
\end{figure}

The focus of the our work here is another, quintessentially quantum way, of exploiting the MAD path feature of quantum networks - that is by combining the end-to-end entanglement established independently along MAD paths between a S-D pair using the fundamental quantum primitive of entanglement purification - to obtain a single effective quantum state of higher fidelity between the given node pair \cite{bennett1996purification,deutsch1996quantum}. We term this strategy multipath entanglement purification (MP-EP). For the QKD example, MP-EP using MAD paths can be advantageous in the situation where each separate path is only able to provide entangled states with entanglement below the DI-QKD threshold \cite{zhang2022device,zapatero2023advances,wehner2018quantum} but when post-purification the fidelity of the state crosses the threshold leading to a non-zero key generation rate.

To evaluate the advantage of MP-EP using MAD paths for entanglement distribution we compare the effective fidelities that this strategy can provide with the single path fidelities achievable using a basic entanglement distribution protocol comprising of only entanglement generation along network edges and swapping at the network nodes. The restriction to this simple basic protocol is not necessary and the method of our comparison can be extended to any distribution protocol utilising single paths between S,D node pairs. The relevant question being: When and what advantage can be achieved by combining the entanglement resources provided by {\it multiple} MAD paths compared to the fidelity of the entanglement connection provided by a {\it single} path between the S,D nodes?

The advantageous scenarios for the MP-EP strategy can be seen statistically by capturing the phenomenology of the probabilistic entanglement manipulation processes in the quantum network. We do so by utilising a statistical model that describes a quantum network using the distribution of its edge-parameters. In this model, to every edge of the network we assign a two-vector whose entries denote the fidelity of the entangled state and the probability of entanglement generation over the edge. The set of parameters for all edges along a given network path yields the two path-parameters, path-fidelity and path-probability, for the basic entanglement distribution protocol. The probability distribution functions of these path-parameters as functions of path length are then obtained from the distributions of the edge-parameters and the former can be used to obtain the average effective fidelity using MP-EP over MAD paths between a pair of S,D nodes as well as the average waiting time for the paths to be available for purification. 

Such an analysis reveals that for appropriately spaced S,D pairs, which depends on the distribution of edge-parameters and topology of the network, MP-EP provides higher average effective fidelity of the entanglement connection relative to entanglement distribution over single paths, on an average over S,D pairs at the same separation. This means that a larger fraction of network node-pairs can have entanglement connections potentially above some desired fidelity threshold, thereby, showing the utility of the MP-EP strategy to increase the functionality of the quantum network. 

We find that MP-EP can successfully boost the effective fidelity between S,D node pairs that have MAD paths that are not too different in their graph lengths, that is, with lengths comparable to the shortest graph distances between the node pair. When the spacing betwen the S,D pair is very short or very large relative to the average length of entangled network paths, MP-EP does not provide any advantage relative to entanglement distribution over single paths. In the region of intermediate spacing between S,D, comparable to the average length of entangled network paths, MP-EP outperforms single path based entanglement distribution yielding higher effective average fidelities between the node pairs.

The structure of this paper is as follows. In the background section, Sec. (\ref{sec:background}), we describe the statistical model of quantum network used in this work in Subsec. (\ref{subsec:Stat_model_QN}), the quantum network operations in Subsec. (\ref{subsec:Q_operations}) and the entanglement distribution protocols in Subsec. (\ref{subsec:Ent_dist}). Next, in Sec. (\ref{sec:stat_network}), We analyse the statistical properties of path parameters by obtaining their distribution functions in Subsec. (\ref{subsec:random_path_parameters}), their mean and standard deviations in Subsec. (\ref{subsec:mean}) and the average length of entangled paths in Subsec. (\ref{subsec:average_length}). In Sec. (\ref{sec:MAD}), first, we mathematically define the set of MAD paths for a pair of source-destination nodes in a network followed by a derivation of the fidelity window for useful entanglement purification in Subsec. (\ref{subsec:fidelity_window}), we then present the statistical criteria for useful MP-EP in Subsec. (\ref{subsec:criteria}) and show the numerical validation of such criteria in Subsec. (\ref{subsec:numerical}). Finally, we conclude with a discussion in Sec. (\ref{sec:discussion}).

{\it Note on terminology -} We use the phrase `fidelity of the path' interchangeably with `fidelity of the state distributed along the path'. Similarly, `purification of the two paths' interchangeably with `purification of the state distributed along the two paths'.

\section{Background}
\label{sec:background}
\subsection{Statistical model of a quantum network}
\label{subsec:Stat_model_QN}
A quantum network (QN) represents a collection of interconnected nodes that are capable of storing, manipulating and transferring quantum information. The edges interconnecting the nodes allow the probabilisitic generation of a shared entangled quantum state, $\rho$, between the two end-nodes of the edge. We assume that successful entanglement generation along an edge, with probability $p$, yields a bipartite entangled state of the isotropic form,
\begin{equation}
    \rho(f)=\frac{(4f-1)}{3}\ket{\phi^+}\bra{\phi^+}+\frac{1-f}{3}\mathbb{1},
    \label{eq:network_state}
\end{equation}
which represents a one-parameter family of mixed states of two-qubits that is widely used in models of quantum networks \cite{mondal2023entanglement,contreras2022asymptotic,santra2019quantum}. Physically, the form of, $\rho$, in Eq. (\ref{eq:network_state}) captures the idea that due to noise in the network channels shared entangled states can be a mixture of a maximally entangled state $|\phi^+\rangle=\frac{1}{\sqrt{2}}(\ket{00}+\ket{11})$ with an unentangled state, the totally mixed state $\mathbb{1}$ of the two qubits. The fidelity, $f:=\langle\phi^+|\rho|\phi^+\rangle$ of the state $\rho$, with, $0\leq f\leq 1$, is directly related to its measure of quantum entanglement. For, $f\in[0.5,1]$, the state $\rho$ has a concurrence \cite{zhu2017operational}, $c(\rho)=(2f-1)$, whereas, for $f\in [0,0.5)$ the state $\rho$ is separable with concurrence equal to zero. Since we are interested in entanglement distribution over network paths potentially comprised of multiple edges we focus on networks where the fidelities of generated states are distributed between $0.5$ and 1. 

Mathematically, a quantum network, can then be represented based on an underlying graph, $\{G(V,E)$, with, $V=\{v_1,v_2,...,v_N\}$, representing the set of nodes and, $E=\{(v_1,v_2),(v_2,v_5),...,(v_i,v_j)\}$, the set of edges which are connected by physical channels such a fiber-optic \cite{brekenfeld2020quantum}, or free-space channels \cite{wehner2018quantum}. Any node $v_i,i=1,...,N$ is assumed to have at least as many qubits as its graph-degree (number of edges connecting to the node). This allows the node to participate in bipartite entanglement generation resulting in states of the form, $\rho$, given by Eq. (\ref{eq:network_state}) with all its neighbors.

The fidelity, $f$, of the state generated over any network edge and the probability, $p$, with which the entanglement generation attempt succeeds are, in general, distinct for each edge of the network. Therefore, for statistically analysing the network as a whole, these edge-parameters, $f,p$, can be considered to be the values realised by the random variables, $F,P$, described by their probability distribution functions, $q_F(f)$ with $0.5\leq f\leq 1$ for the fidelity, and, $q_P(p)$ with $p_{\text{min}}\leq p\leq 1$ for the probability.

Of interest, for any entanglement distribution protocol, such as the one we describe in Subsec. (\ref{subsec:Ent_dist}), are the probability distribution functions of the path-parameters obtained along network paths comprised of multiple network-edges. For a path comprised of $l$-edges, with $l=2,3,...$, we use the notations, $q_{F}^{(l)}(f^{(l)})$ and $q_{P}^{(l)}(p^{(l)})$, respectively, for the probability distribution functions (pdf) of the fidelity and probability of the end-to-end state obtained using the given entanglement distribution protocol. For, $l=1$, these distribution functions, of course, are the edge-parameter distributions for which we omit the superscipt notation and simply use $q_F(f)$ and $q_P(p)$ to represent the p.d.f. of edge fidelity $(f)$ and edge probability $(p)$ respectively. For, $l>1$, the distribution functions $q_{F,P}^{(l)}(f^{(l)})$ are obtained as appropriate convolutions of the edge-parameter distributions, $q_{F,P}(f)$, as we show in Sec. (\ref{sec:stat_network}).

\subsection{Quantum network operations}
\label{subsec:Q_operations}
There are three quantum network operations relevant for the entanglement distribution protocol we consider in this paper, namely (i) entanglement generation, (ii) entanglement swapping, and (iii) entanglement purification. For a network path comprised of $l$-edges, a sequence of these operations on the edges and nodes along the path enables the distribution of end-to-end entanglement. In the following, we discuss these operations briefly and connect them to the parameters of the statistical model of the quantum network described above.

\subsubsection{Entanglement Generation}
\label{subsubsec:Ent_gen}

Entanglement generation over the network edges probabilistically and asynchronously creates entangled edges in the network with probability, $p$, that can vary over different edges of the network with an overall probability distribution function given by $q_P(p)$. The generated entangled state is of the form, $\rho(f)$ given by Eq. (\ref{eq:network_state}), with a fidelity, $f$, that can also vary over the different edges of the network with its overall distribution function given by $q_F(f)$. Since this entanglement is between two qubits at  nearest-neighbor nodes according to the graph topology, $G(V,E)$, we term this entanglement short-ranged. 

The actual physical process of entanglement generation involves a source of entangled pairs of photons \cite{zhang2021high,barz2010heralded,wang2024uncovering} and the transmission of both photons in the pair to the opposite end-nodes of an edge. The photons arriving at the nodes may be stored in quantum memories such as those based on atomic ensembles \cite{reim2011single} or rare -earth ion-doped crystals \cite{tittel2025quantumnetworksusingrareearth} for further entanglement manipulations. The fidelity of the generated entangled state, $f$, effectively captures the decoherence in the quantum channel connecting the source of entangled photons to the end-nodes as well as the decoherence in the quantum memories. The probability, $p$, effective accounts for the probability of the intrinsically probabilistic process of generating entangled-photon pairs \cite{zhang2021high,barz2010heralded,wang2024uncovering} as well as the losses in the channels \cite{takeoka2014fundamental}.

\subsubsection{Entanglement  Swapping}
\label{subsubsec:Ent_swap}

Entanglement swapping at the nodes of the network connects the entangled states generated over two adjacent edges of the node. This leads to an entangled state shared by nodes which were not nearest neighbors according to the topology of the network graph, $G(V,E)$, allowing long-range entanglement distribution. By performing entanglement swapping operations at every intermediate node along a network path, end-to-end entanglement can be distributed between any pair of S-D nodes.

Entanglement swapping is performed by measuring two qubits at a node, one from each of two adjoining entangled states, in the complete basis of maximally entangled pure states of the two qubits \cite{zukowski1993event,pan1998experimental} such as the Bell basis spanned by $\{\ket{\psi^{\pm}}:=(\ket{01}\pm\ket{10})\sqrt{2},\ket{\phi^{\pm}}:=(\ket{00}\pm\ket{11})\sqrt{2}\}$. For two states, $\rho_{v_1v_2}=\rho(f_1)$ and $\rho_{v_2v_3}=\rho(f_2)$ with fidelities $f_1$ and $f_2$ over the edges $(v_1,v_2)$ and $(v_2,v_3)$, respectively, the fidelity of the entangled state obtained after swapping depends on the input fidelities, $f_1$ and $f_2$. Measurement of one qubit from each of the two states at the common node, $v_2$, yields one of the four states with equal probability ($1/4$), $\rho_{out}=\{\rho_{v_1v_3}^{\psi^+},\rho_{v_1v_3}^{\psi^-},\rho_{v_1v_3}^{\phi^+},\rho_{v_1v_3}^{\phi^-}\}$, between nodes, $v_1$ and $v_3$, depending on the Bell state measurement outcome. All four states in the set $\rho_{out}$ can be transformed in to each other via local unitary operations and have the common form, $\rho(f')$ given by Eq. (\ref{eq:network_state}), with fidelity, $f'=(1/4)+(3/4)\{(4f_1-1)/3\}\{(4f_2-1)/3\}$. 

For a network path comprised of $l$-edges with states $\rho(f_i)$ of fidelities $f_1,f_2,...,f_l$ along the edges, entanglement swapping at the intermediate nodes yields a state $\rho(f^{(l)})$ of the same form as in Eq. (\ref{eq:network_state}) with the fidelity given by,
\begin{align}
    f^{(l)}&=\frac{1}{4}+\frac{3}{4}\prod_{i=1}^l\bigg(\frac{4f _{i}-1}{3}\bigg).
    \label{eq:swapping_fidelity}
\end{align}
This output fidelity characterises the end-to-end entangled state obtained along the path and is independent of the order of the states,  $\rho(f_i)$, over the edges along the path for states of the form Eq. (\ref{eq:network_state}). The output path fidelity is upper bounded by the least of the edge fidelities along the path. 

Finally, note that the output fidelity $f^{(l)}$ can be understood as a random variable that is a function of the random fidelities, $f_i$, of the states along the network edges. All edge fidelities are assumed to be chosen from the independent and identical distributions (i.i.d) of the edge fidelity, $q_F(f)$. Therefore, the distribution, $q_{F}^{(l)}(f^{(l)})$, of $f^{(l)}$ can be obtained using the $l$-fold convolution of $q_F(f)$.

The probability, $p^{(l)}$, of obtaining an end to end state is given simply by the product of the edge probabilities, 
\begin{align}
p^{(l)}=\prod_{i=1}^lp_i,
\label{eq:path_prob}
\end{align}
where each of the edge probabilities are independent and identically distributed according to the function, $q_P(p_i)$.

\subsubsection{Entanglement Purification}
\label{subsubsec:Ent_pur}
Entanglement purification over network edges or paths connecting two network nodes, S,D, probabilistically improves the fidelity of the shared entangled state between the nodes - that is, it can be used to strengthen the entanglement connection between the said nodes. For two states, $\rho(f_1)$ and $\rho(f_2)$ of the form Eq. (\ref{eq:network_state}), Deutsch's protocol for purification \cite{deutsch1996quantum} uses the qubits of one pair as the control and the other pair as targets to apply bilateral CNOT gates at nodes S and D. The target qubits are then measured in the computational basis and the local results compared via classical communication. In case of concurrent measurement results (both $0$ or both $1$) the pair of qubits used as control will be found in a diagonal state in the Bell basis with a fidelity,
\begin{align}
     &f_{\rm out}(f_1,f_2)=\nonumber\\
     &\frac{f_1f_2+\frac{1}{9}(1-f_1)(1-f_2)}{f_1f_2+{\frac{1}{3}(f_1(1-f_2)+(1-f_1)f_2})+\frac{5}{9}(1-f_1)(1-f_2)},
     \label{eq:fout}
\end{align}
which occurs with a probability given by the denominator of the right hand side of Eq. (\ref{eq:fout}).
This output fidelity $f_{\rm out}(f_1,f_2)$ is always greater than the input fidelity when the two input fidelities are identical and greater than the purification threshold of $0.5$, that is, $f_1=f_2>0.5$. 

However, when the two input fidelities are not the same, say $f_1>f_2$, the output is greater than the larger of the two, $f_{\rm out}(f_1,f_2)>f_1$, only if $f_2$ is within a certain range of values determined by $f_1$ - a scenario we discuss in detail in Subsec. (\ref{subsec:fidelity_window}). This situation with non-identical input fidelities is crucial to our analysis of MP-EP along MAD paths because, in general, the fidelities of the states distributed over MAD paths between a pair of end nodes are distinct. 

In this manner, entanglement purification lets one obtain higher fidelity entangled states between a given S,D pair by using lower fidelity entangled states probabilistically but in a heralded manner. The higher fidelity output state can, therefore, be always postselected.


\subsection{Entanglement distribution protocol}
\label{subsec:Ent_dist}
The basic entanglement distribution protocol considered in this work for distributing an entangled quantum state between arbitrary network nodes S,D involves the following steps:
\begin{itemize}
\item Step 1: Identification of a network path, $R:=\{(S,v_{i_1}),(v_{i_1},v_{i_2}),(v_{i_2},v_{i_3}),....,(v_{i_{l-1}},D)\}$, between the S,D nodes comprising a sequence of $l$-adjacent nodes.
\item Step 2: Probabilistic entanglement generation over all $l$-edges resulting in states $\rho(f_i)$ of the form Eq. (\ref{eq:network_state}) over each edge with the random fidelities $f_i$ distributed according to the distribution function, $q_F(f_i)$.
\item Step 3: Entanglement swapping at all intermediate nodes, $v_{i_1},v_{i_2},...,v_{i_{l-1}}$, along the path, $R$, yielding an end-to-end state between the nodes, S,D, also of the form $\rho(f^{(l)})$ given by Eq. (\ref{eq:network_state}) and fidelity $f^{(l)}$ given by Eq. (\ref{eq:swapping_fidelity}).
\end{itemize}

The above protocol is probabilistic and heralded since each successful entanglement generation attempt over the edges along the path is heralded. The network path $R$ that connects the source and destination nodes, S,D, is chosen to be the shortest graph path (SGP) which can be determined using the Dijkstra's algorithm \cite{chao2010developed} given the network graph, $G(V,E)$. The entanglement swapping operations in step 3 can occur whenever two adjacent edges of a node along the path, R, have heralded successful entanglement generation. Therefore, steps 2 and 3 may be implemented in an interleaved manner rather than in a fully chronologically sequential manner. 

In addition to the above basic entanglement distribution protocol, the entanglement connection between S,D nodes can be optionally strengthened using multipath entanglement purification when two or more alternative, distinct paths $R_1,R_2$ etc. are available between them and fidelities of the states provided by the two paths are such that the resultant output fidelity after purification is greater than either of the two input fidelities as described in Eq. (\ref{eq:fout}). When this optional advanced protocol of MP-EP using MAD paths is to be implemented, step 1 of the basic protocol above needs to be modified to identify at least two MAD paths between the S,D nodes. This can be achieved by sequential application of the Dijkstra's algorithm as shown in the Appendix (\ref{app:algorithm}). This modified step 1 is then followed by application of steps 2 and 3 independently along the available MAD paths. Further, the following additional step is required,
\begin{itemize}
\item Step 4: Entanglement purification of two states distributed between S,D using paths $R_1$ and $R_2$.
\end{itemize}

Our results in the following sections of this paper will show that this optional advanced entanglement distribution protocol utilising MP-EP along MAD paths is useful only for S, D node pairs separated over certain ranges of the graph distance. This region of graph distances is determined by the statistical features of the distributions of the edge parameters, $q_F(f),q_P(p)$.

\section{Statistical analysis of network path parameters}
\label{sec:stat_network}

Here we derive the probability distribution function of the path parameters, $q_F^{(l)}(f^{(l)}),q_P^{(l)}(p^{(l)})$, obtained as a result of applying the basic entanglement distribution protocol described above. An analysis of these distribution functions reveals the regions of the network graph where MP-EP along MAD paths can be usefully deployed.

\subsection{Random path-parameters in a quantum network}
\label{subsec:random_path_parameters}

\begin{figure}
    \centering
    \includegraphics[width=\linewidth]{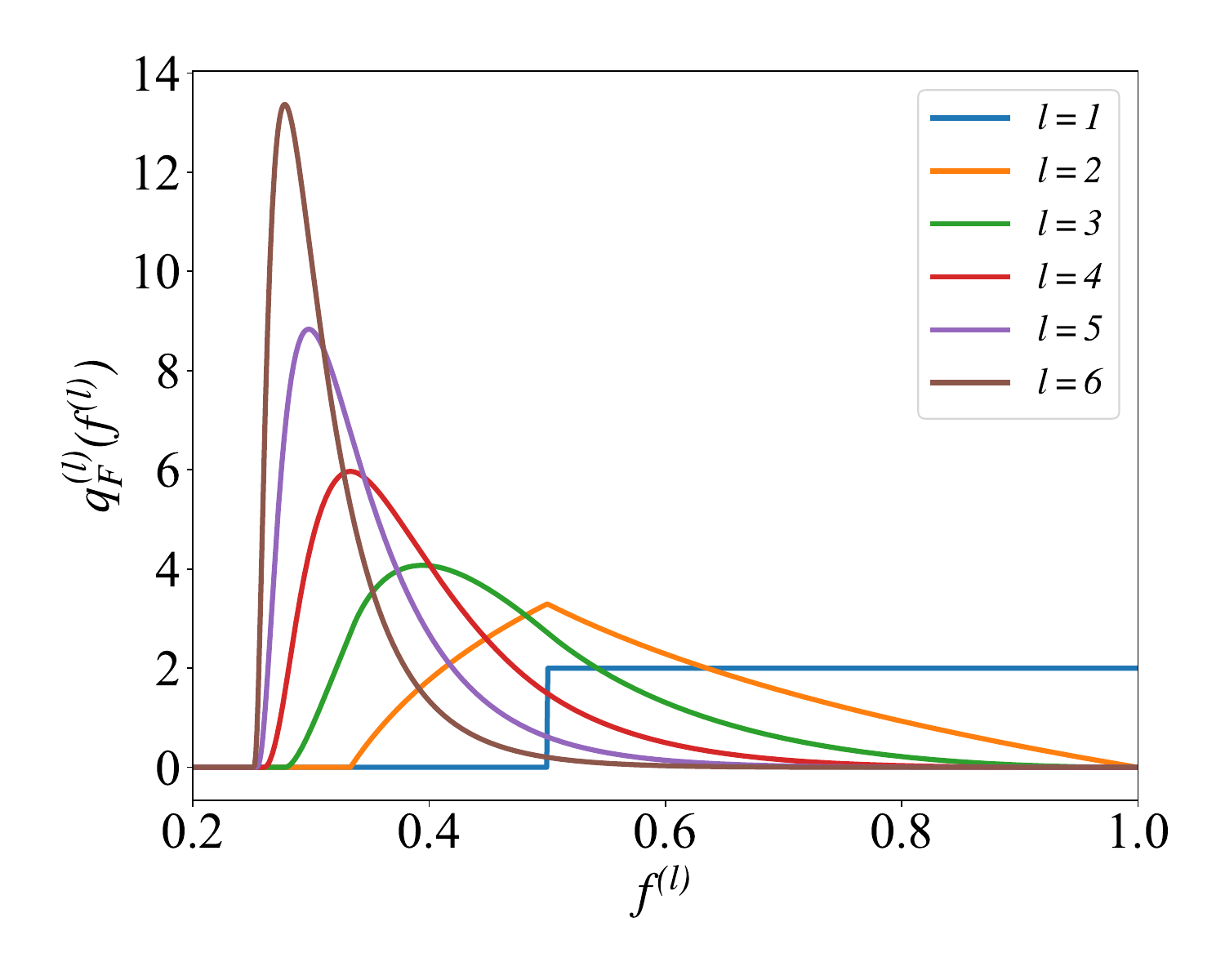}
    \caption{The p.d.f., $q_F^{(l)}(f^{(l)})$ Eq. (\ref{eq:fidelity_pdf}), of path fidelity, $f^{(l)}$, for uniform edge fidelity distribution, $U(0.5,1)$, for different path-lengths $l\in[1,2,\cdots,6]$. The shifting of maximum probability weight towards lower fidelity with increasing $l$ indicates the decrement in the path fidelity with more number of entanglement swapping.
    }
    \label{fig:fidelity0.5}
\end{figure}
For a network path comprised of $l$-edges between a pair of S, D-nodes with the states, $\rho(f_i)$ of the form Eq. (\ref{eq:network_state}), with fidelities, $f_{\rm min}\leq f_i\leq 1,i=1,2,...,l$ along the edges, the end-to-end fidelity, $f^{(l)}$, is given by Eq. (\ref{eq:swapping_fidelity}). This implies that,
\begin{equation}
    \log{\Big(\frac{4f^{(l)}-1}{3}\Big)}=\sum_{i=1}^l \log{\Big(\frac{4f_i-1}{3}\Big)},
\end{equation}
from which the p.d.f. of the random variable $\log((4f^{(l)}-1)/3)$ can then be obtained as the $l$-fold convolution of the p.d.f. of the transformed edge variable $\log((4f-1)/3)$. The p.d.f. of the random path fidelity, $q_F^{(l)}(f^{(l)})$, is then obtained by transforming the above obtained p.d.f. for $\log((4f^{(l)}-1)/3)$ back.

For the important, special case of uniformly distributed edge fidelities between $[f_{\rm min} ,1]$, that is, $q_F(f )\sim U(f_{\rm min} ,1)$ we obtain the p.d.f. of path fidelity as, 
\begin{align}
    q_F^{(l)}(f^{(l)})=&\Big(\frac{3}{4}\Big)^{l-1}\frac{1}{(l-1)!}\Big(\frac{1}{1-f _{\rm min}}\Big)^l\nonumber\\
    &\sum_{n=0}^{(m-1)}(-1)^n\binom{l}{n}\Bigg[\log{\bigg(\frac{(4f_{\rm min} -1)^n}{(4f^{(l)}-1)3^{n-1}}\bigg)}\Bigg]^{l-1},
    \label{eq:fidelity_pdf}
\end{align}
where, the random path fidelity $f^{(l)}$ can lie in one of the `$l$' geometrically-spaced intervals of path fidelity, 
\begin{align}
\frac{1}{4}+\frac{3}{4}\Big(\frac{4f_{\rm min} -1}{3}\Big)^m\leq f^{(l)}\leq \frac{1}{4}+\frac{3}{4}\Big(\frac{4f_{\rm min} -1}{3}\Big)^{m-1},
\label{fid_interval}
\end{align}
with, $m=1,2,...,l$, between a minimum of $f_{\rm min}^{(l)}=(1/4)+(3/4)((4f_{\rm min} -1)/3)^l$ and the maximum of $f_{\rm max}^{(l)}=1$.
The summation over the index `$n$' in the r.h.s. of the Eq. (\ref{eq:fidelity_pdf}) running from $n=0,1,...,(m-1)$, therefore, provides the distinct functional forms of the p.d.f. for the $l$-different intervals. The p.d.f. $q_F^{(l)}(f^{(l)})$ over the allowed ranges of $f^{(l)}$ for different values of path length, $l={1,2,...,6}$, for uniformly distributed edge fidelities, $q_F(f )\sim U(0.5,1)$, is shown in Fig. \ref{fig:fidelity0.5}. We make several important observations regarding the p.d.f. $q_F^{(l)}(f^{(l)})$ in the following paragraphs while its complete derivation is provided in Appendix (\ref{app:pdf_path_fidelity}).

First, notice from Fig. \ref{fig:fidelity0.5} that the peak of the p.d.f. shifts towards the minimum possible fidelity $f_{\rm min}^{(l)}$ for a given path length with increasing values of $l$. Moreover, with increasing $l$ the p.d.f. gets tightly concentrated around the value of $f^{(l)}$ corresponding to the peak of the distribution. This is consistent with Eq. (\ref{eq:swapping_fidelity}) and reflects the fact that entanglement swapping of states over edges of progressively longer paths yield end-to-end states with decreasing fidelity.
 
Second, notice from the same figure that the cumulative probability for path fidelities, $\int_{0.5}^1{\rm d}f^{(l)}q_F^{(l)}(f^{(l)})$, in the entangled region of fidelity, $0.5<f^{(l)}\leq 1$, decreases with increasing path length, $l$.  An important implication is that for long network paths the path fidelity is statistically unlikely to be above the entanglement threshold fidelity.

\begin{figure}
    \centering
    \includegraphics[width=\linewidth]{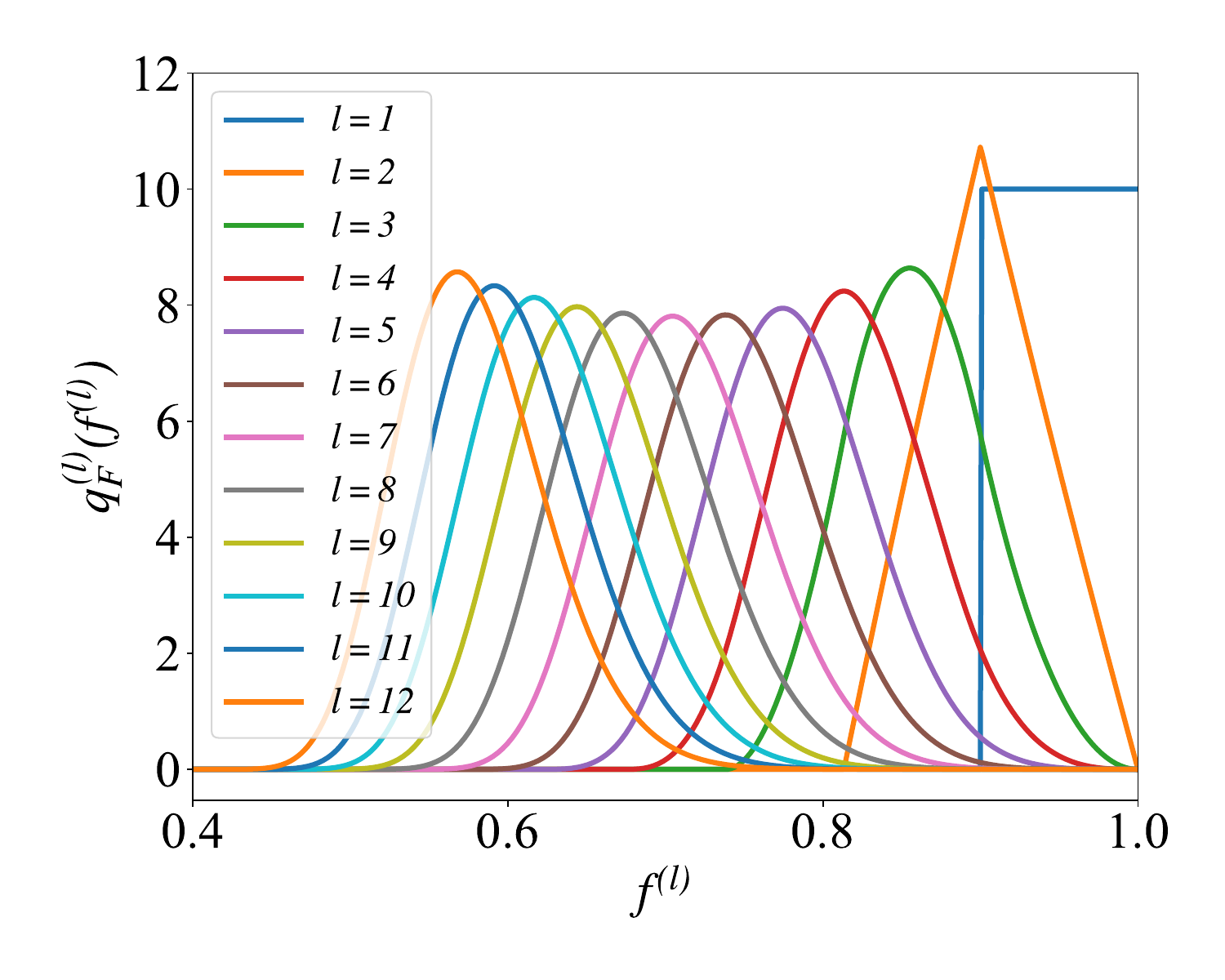}
    \caption{The p.d.f., $q_F^{(l)}(f^{(l)})$ Eq. (\ref{eq:fidelity_pdf}), of path fidelity, $f^{(l)}$, for narrower edge fidelity distribution $U(0.9,1)$, for path-lengths $l\in\{1,2,\cdots,12\}$. Though the path-fidelity decreases with increasing $l$, but states remain entangled for sufficiently high $l$. This indicates that with narrower edge fidelity distributions, end-to-end communication can be realized for large $l$. }
    \label{fig:fidelity0.9}
\end{figure}

Third, we remark that the value of the p.d.f. $q_F^{(l)}(f^{(l)})$ for a given fidelity, $f^{(l)}$, increases with the minimum edge fidelity, $f _{\rm min}$, for a given path length $l$ - as can be understood by analysing the r.h.s of Eq. (\ref{eq:fidelity_pdf}). Since for the uniform distribution, the average edge fidelity, $\overline{f }=(1+f _{\rm min})/2$, also increases with the minimum fidelity, networks with higher mean edge fidelities have path fidelity distributions with fatter tails towards the maximum $f^{(l)}=1$ for a given $l$. Thus, for networks with high mean edge fidelities relatively longer network paths can have significant cumulative probabilities, $\int_{0.5}^1{\rm d}f^{(l)}q_F^{(l)}(f^{(l)})$, for the path fidelities in the entangled region, $0.5<f^{(l)}\leq 1$. This can be seen by comparing Figs. \ref{fig:fidelity0.5} and \ref{fig:fidelity0.9} which consider two networks with $q_F(f )\sim U(0.5,1)$ and $q_F(f )\sim U(0.9,1)$ respectively. In the first case, from Fig. \ref{fig:fidelity0.5}, we see that paths with $l\geq 6$ have a negligible probability to have fidelities above $f^{(l)}=0.5$, whereas, in the second case, Fig. \ref{fig:fidelity0.9}, much longer paths upto $l=12$ have almost their entire cumulative probability above that value of fidelity.

The p.d.f. of the path probability, $q_P^{(l)}(p^{(l)})$, can be derived using arguments similar to the ones above for the path fidelities. Taking the logarithm of both sides of Eq. (\ref{eq:path_prob}) one gets,
\begin{align}
\log p^{(l)}=\sum_{i=1}^{i=l}\log p _i.
\end{align}
Thus, the $l$-fold convolution of the p.d.f. of the transformed variable, $\log p _i$, yields that for $\log p^{(l)}$ from which the p.d.f. of the path probability can be obtained after a transformation of variables.

\begin{figure}
    \centering
\includegraphics[width=\linewidth]{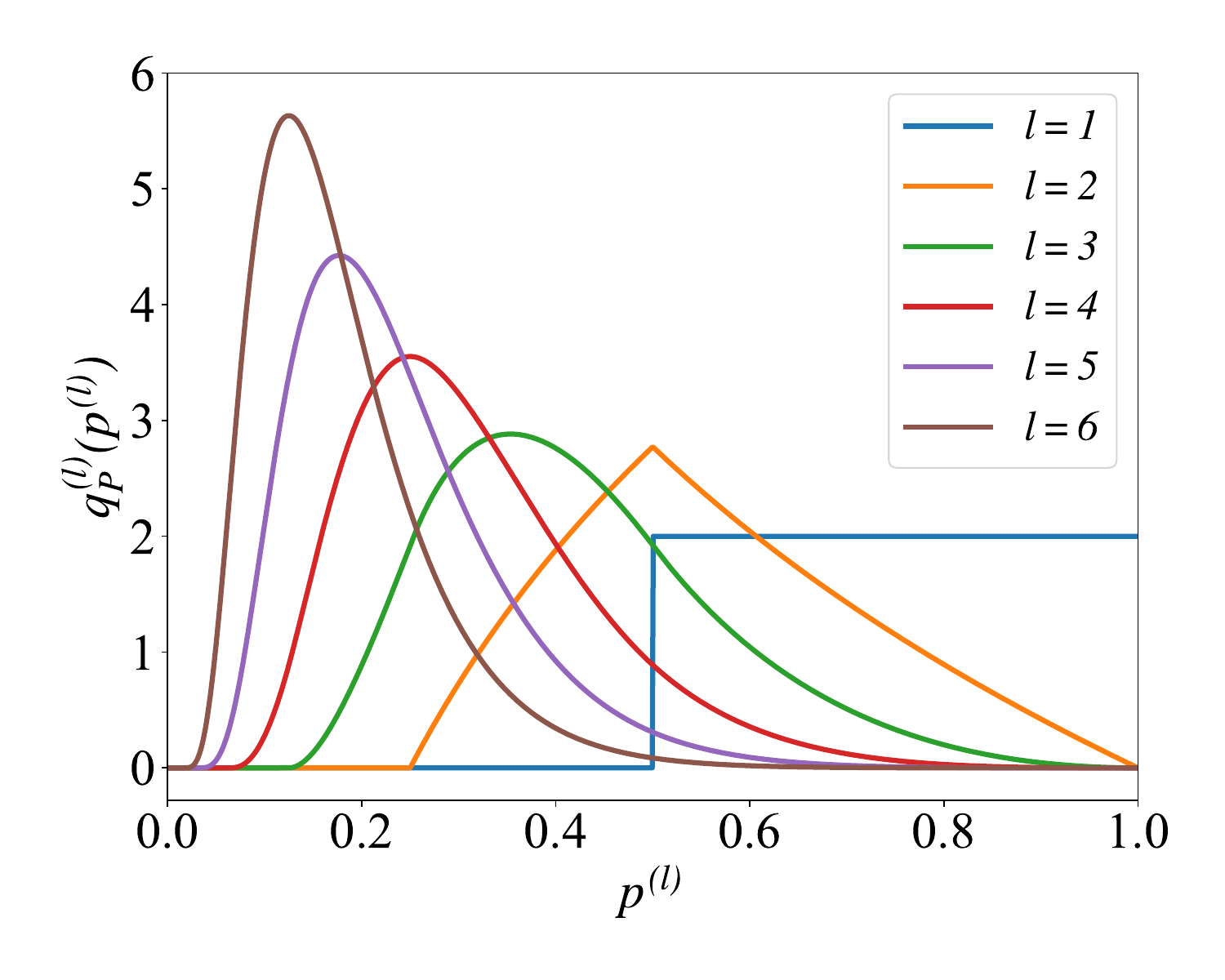}
    \caption{The p.d.f. of path probability, $q_P^{(l)}(p^{(l)})$ Eq. (\ref{eq:prob_pdf}) w.r.t path probability $p^{(l)}$ when edge probabilities are distributed following the distribution $U(0.5,1)$ for path-lengths $l\in\{1,2,\cdots,6\}$. The p.d.f. shifts towards lower probability values with increasing $l$ indicating that longer paths would be available in longer times.}
    \label{fig:prob_pdf}
\end{figure}

For the important, special case of uniformly distributed edge probabilities between, $[p_{\rm min} ,1]$, that is, $q_P(p )\sim U(p_{\rm min} ,1)$, we obtain the path probability distribution as,
\begin{align}\label{eq:prob_pdf}
      q_P^{(l)}(p^{(l)})=&\frac{1}{(l-1)!}\Big(\frac{1}{1-p _{\rm min}}\Big)^l\nonumber\\
      &\sum_{n=0}^{(m-1)}(-1)^n\binom{l}{n}\Big(\log\Big[\frac{(p_{\rm min} )^n}{p^{(l)}}\Big]\Big)^{l-1},
\end{align}
where the path probability can lie in one of the $l$ geometrically spaced intervals, $(p_{\rm min} )^m\leq p^{(l)}\leq (p_{\rm min} )^{m-1}$ with $m=1,2,\cdots,l$. The minimum path fidelity for a path of length $l$ is, therefore, $p^{(l)}=(p_{\rm min} )^l$ while the maximum is, $p^{(l)}_{\rm max}=1$. The p.d.f. $q_P^{(l)}(p^{(l)})$ for different values of the path length, $l=1,2,...,6$, for uniformly distributed edge probabilities, $q_P (p )\sim U(0.5,1)$, is shown in Fig. \ref{fig:prob_pdf}.

The p.d.f. of the path probabilities, $q_P^{(l)}(p^{(l)})$, shows the same qualitative behavior as that for the path fidelities as can be seen by comparing Fig. \ref{fig:prob_pdf} with Figs. \ref{fig:fidelity0.5} and \ref{fig:fidelity0.9}. Essentially, the p.d.f. of the path probabilities gets progressively squeezed towards the minimum possible path probability with increasing path length $l$. Increasing the mean edge probability one gets path probability distributions with fatter tails extending to the maximum path fidelity, $p^{(l)}_{\rm max}=1$. Therefore, for networks with higher mean edge probabilities longer paths can have higher cumulative probabilities of paths above a given threshold probability.

\subsection{Mean and standard deviation of random path parameters}
\label{subsec:mean}
The average values and standard deviations of the random path fidelity, $f^{(l)}$, and path probability, $p^{(l)}$, for paths of a given length, $l$, can be readily calculated from the p.d.f.s of the path parameters given by Eqs. (\ref{eq:fidelity_pdf}) and (\ref{eq:prob_pdf}) or more directly from Eqs. (\ref{eq:swapping_fidelity}), (\ref{eq:path_prob}). These statistical moments of the path provide intuition about the scaling of the path parameters for varying separation between the source, destination nodes in the network.

The statistical mean of the random path fidelity and its standard deviation for a path of length $l$ evaluate to,
\begin{align}
    \overline{f^{(l)}}&=\frac{1}{4}+\frac{3}{4}\bigg(\frac{4\overline{f }-1}{3}\bigg)^l,\label{eq:mean_path_fidelity}\\
    \sigma_{f^{(l)}}&=\frac{3}{4}\Bigg(\left[\left(\frac{4\overline{f }-1}{3}\right)^2+\frac{16}{9}\sigma_{f }^2\right]^l -\left(\frac{4\overline{f }-1}{3}\right)^{2l}\Bigg)^{0.5},\label{eq:std_dev_path_fidelity}
\end{align}
where, on the r.h.s. of the equations above $\overline{f }, \sigma_{f }$ are the mean and standard deviation of the edge fidelity. 

Similary, the statistical mean and standard deviation of the path probability are obtained as,
\begin{align}
\overline{p^{(l)}}&=\overline{p}^l,\\
\sigma_{p^{(l)}}&=((\overline{p }^2+\sigma_{p }^2)^l-\overline{p }^{2l})^{0.5},
\label{eq:mean_path_prob}
\end{align}
where, $\overline{p }, \sigma_{p }$ are the mean and standard deviation of the edge probability. 

\begin{figure}
    \centering
    \includegraphics[width=\linewidth]{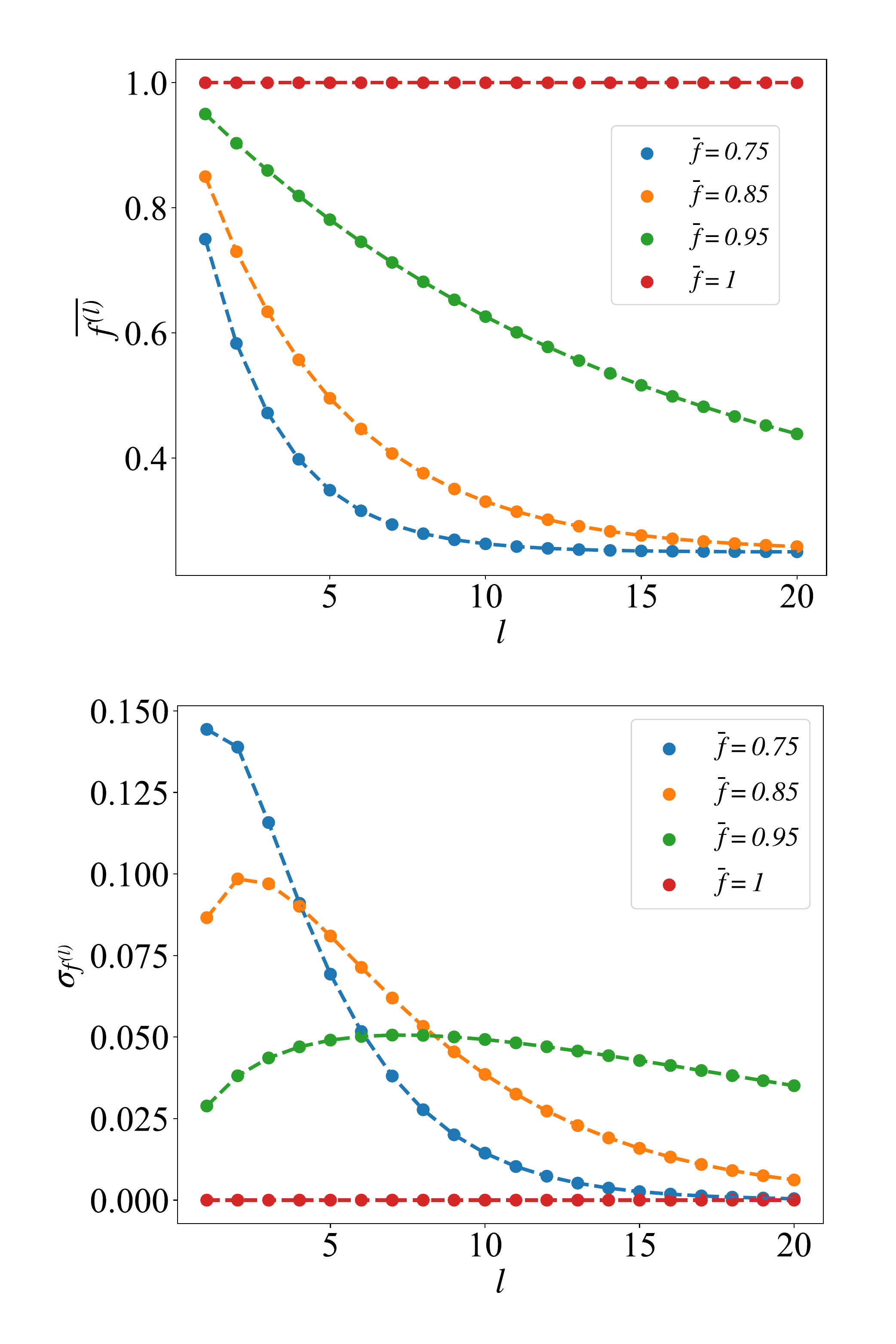}
    \caption{The average path fidelity, $\overline{f^{(l)}}$, (top panel) and its standard deviation, $\sigma_{f^{(l)}}$, (bottom panel) vs the path length, $l$, for uniformly distributed edge fidelities for four different average edge fidelities, $\overline{f }= 0.75 {~\rm (Blue)},~0.85 {~\rm (Orange)}, ~0.95 {~\rm (Green)}~\text{and}~1 {~\rm (Red)}$. The actual data points are obtained for positive integer values of the path lengths but the line connecting them is used to indicate the scaling.}
    \label{fig:std_fl_scaling} 
\end{figure}

While both the average path fidelity and path probability, $\overline{f^{(l)}},\overline{p^{(l)}}$, monotonically decrease with the length $l$ - the standard deviations of these path parameters $\sigma_{f^{(l)}},\sigma_{p^{(l)}}$ can show interesting non-monotonic behavior with the path length, see Fig. \ref{fig:std_fl_scaling}. It is clear from the figure the standard deviation of the path fidelity can actually increase with $l$ for certain ranges of mean edge fidelities. However, for very long paths both the mean path fidelity and mean path probability approaches the absolute minimum fidelity, $\overline{f^{(l)}}\to 1/4$, and absolute minimum probability of $0$ respectively. For such long paths the standard deviations of both parameters becomes negligible due to concentration of measure. This can be clearly seen in the situation when narrow edge parameter distributions are considered, that is, those with $\sigma_{f^{(l)}}\ll \overline{f^{(l)}}$ and $\sigma_{p^{(l)}}\ll \overline{p^{(l)}}$ in which case the standard deviations obtained from Eqs. (\ref{eq:mean_path_fidelity}) and (\ref{eq:mean_path_prob}) scale as,
\begin{align}
\sigma_{f^{(l)}}&\sim \sqrt{l}\left(\frac{4\overline{f }-1}{3}\right)^{l-1}\sigma_{f },\\
\sigma_{p^{(l)}}&\sim \sqrt{l}\left( \overline{p }\right)^{l-1}\sigma_{p }.
    \label{std_pl_app}
\end{align}

\subsection{Average length of entangled network paths} 
\label{subsec:average_length}
Relevant for purposes of entanglement distribution are network paths with fidelity, $f^{(l)}>0.5$. The average length of such paths in terms of graph distance can be obtained using the expression for average path fidelity, Eq. (\ref{eq:mean_path_fidelity}), yielding,
\begin{align}
    l_{\rm avg}= \Big\lfloor \dfrac{-\log(3)}{\log\Big(\frac{4\overline{f }-1}{3}\Big)}\Big\rfloor.
    \label{eq:l_avg}
\end{align}
Source-destination pairs separated by the graph distance, $l_{\rm avg}$, are on average, able to share entangled quantum states using the basic entanglement distribution protocol. 

The distance, $l_{\rm avg}$, provides a comparative scale to classify network paths based on their length. Any path in the network with length, $l$, up to, $\sim l_{\rm avg}$, can be considered to be a {\it short} network path and the basic protocol can distribute entangled states between S,D nodes connected by such short network paths. The average path fidelities for very short network paths, $l\ll l_{\rm avg}$, is close to the maximum of the edge fidelity distributions, i.e., $\overline{f^{(l)}}\approx \overline{f}$. On the other hand, any path in the network with, $l\gg l_{\rm avg}$, can be considered to be a {\it long} network path over which entangled states cannot be distributed, on average. The average path fidelity for long network paths falls below the threshold for entanglement, i.e., $\overline{f^{(l)}}<0.5$. 

For multipath entanglement distribution long network paths are therefore useless because purification protocols require the path fidelity of the state to be $f^{(l)}>0.5$. Whereas, for very short network paths it does not provide much of an advantage in achieving higher effective fidelity between the S,D pair because the average single path fidelities are already close to the average of the edge fidelities and improving that would require alternate paths of similar short lengths which typical network topologies do not provide. MP-EP is found to be a useful entanglement distribution strategy for S,D pairs separated by shortest graph paths of intermediate lengths, $l\lesssim l_{\rm avg}$, as we discuss next.

\section{Multipath entanglement purification on Multiple, Alternative, Distinct network paths}
\label{sec:MAD}
\begin{figure}
    \centering
    \includegraphics[width=\linewidth]{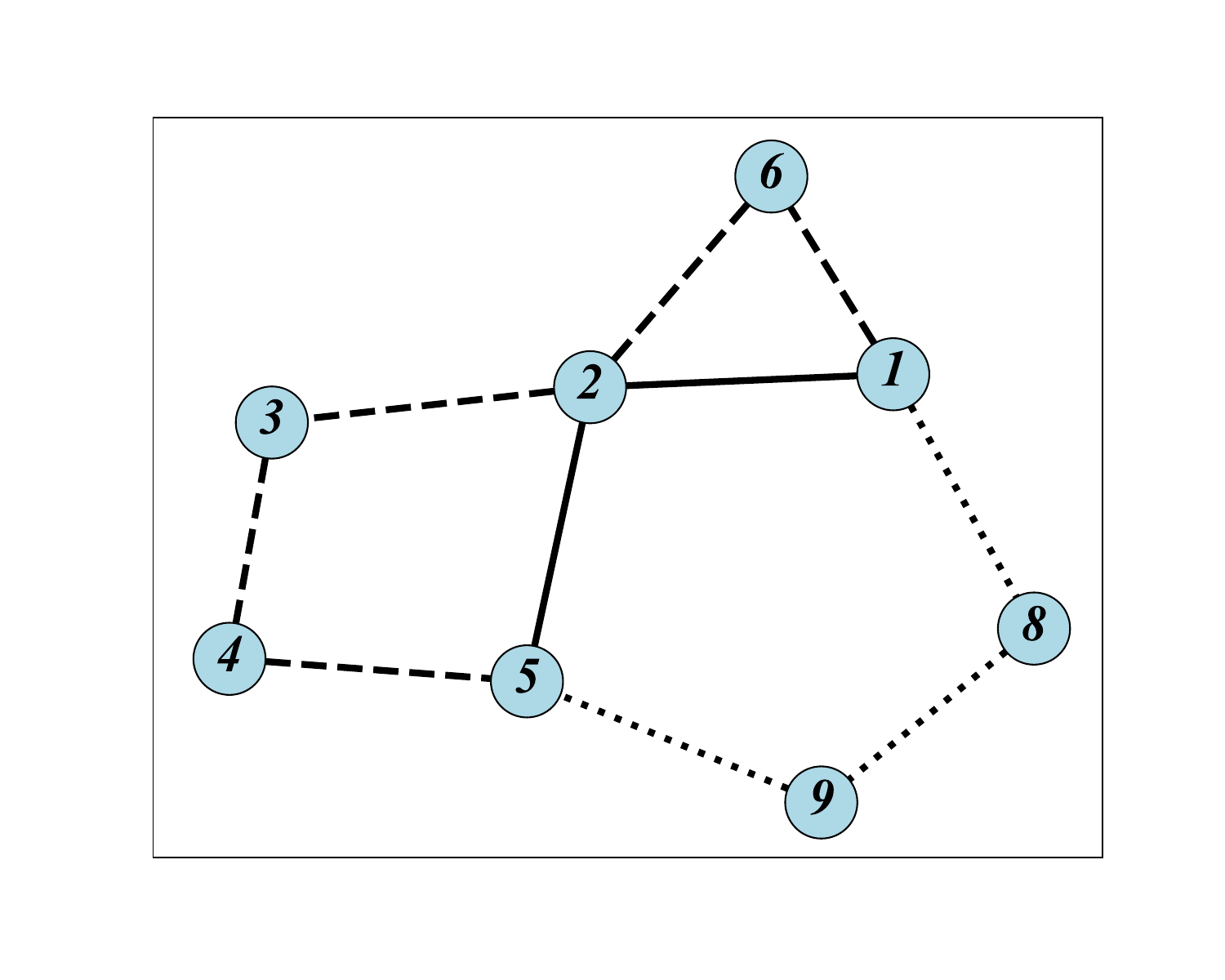}
    \caption{Illustration of MAD paths in a quantum network. Shown is a network of $|V|=9$ nodes interconnected by $|E|=10$ edges with nodes $(1,5)$ as the S,D pair. This pair of nodes can be connected via three MAD paths, ${R}_1=\{(1,2),(2,5)\},~{ R}_2=\{(1,8),(8,9),(9,5)\}~\text{and}~ { R}_3=\{(1,6),(6,2),(2,3),(3,4),(4,5)\}$. Notice that, $R_1$ and $R_3$ share node $2$ as common, while, none of the three paths share an edge in common.}
    \label{fig:MAD_paths}
\end{figure}

Multipath entanglement purification between a pair of S,D nodes can be performed if there exist in the network multiple, alternative, distinct paths connecting the two nodes. The set, $\mathcal{R}$, of $k$ such MAD network paths, $R_1^{S-D},R_2^{S-D},...,R_k^{S-D}$, can be defined as,
{\small
\begin{align}
    \mathcal{R}^{S-D} := \{ \{R_i^{S-D}\}_{i=1}^{i=k} \mid E(R_i^{S-D}) \cap E(R_j^{S-D}) = \emptyset, \ \forall \ i \neq j \}.
    \label{MAD_Paths}
\end{align}}
where, $E(R_i^{S-D})$ represents the set of edges included in the path $R_i^{S-D}$. That is, MAD paths have no common edges though they can intersect at common nodes. The same is depicted in Fig. \ref{fig:MAD_paths}. 

Assuming that the network graph $G(V,E)$ permits a set $\mathcal{R}^{S-D}$ of MAD paths with at least two paths, say $R_1^{S-D},R_2^{S-D}$, the random fidelities, $f_1,f_2$, of the states established independently over these two paths using the basic entanglement distribution protocol need to be sufficiently close to each other in order for the post purified state to have a higher fidelity. Only in this situation is multipath entanglement purification useful. The calculation of this useful purification window is considered in the following.


\subsection{Fidelity window for useful entanglement purification}
\label{subsec:fidelity_window}

Consider entanglement purification of two states, $\rho(f_1),\rho(f_2)$, of the form Eq. (\ref{eq:network_state}) using Deutsch's protocol for entanglement purification. An output state will be a diagonal state in the Bell-basis with fidelity, $f_{\rm out}(f_1,f_2)$, given by Eq. (\ref{eq:fout}) will be obtained probabilistically and this process is useful to increases the fidelity if,
\begin{align}
f_{\rm out}(f_1,f_2)\geq \text{Max}(f_1,f_2).
\label{ineq:purif_cond}
\end{align}
This condition is satisfied if the difference of the random path fidelities, $D(f_1,f_2):=f_1-f_2$, lies within the interval,
\begin{align}
\check{D}(f_1)\leq D(f_1,f_2)\leq \hat{D}(f_1),
\label{ineq:D_up_down}
\end{align}
where, the upper and lower limits, $0\leq\hat{D}(f_1), \check{D}(f_1)\leq 0$, are given by,
\begin{align}
\check{D}(f_1)&=\frac{(8f_1^2-8f_1+3-\sqrt{28f_1^2-26f_1+7})}{(8f_1-2)},\nonumber\\
\hat{D}(f_1)&=\frac{(8 f_1^3-14 f_1^2+7 f_1-1)}{(8f_1^2-12 f_1+1)}.
\label{eq:D_limits}
\end{align}
The fidelity intervals, 
\begin{align}
\Delta_{\text{upper}}(f_1)&:=[f_1,f_1-\check{D}(f_1)],\\
\Delta_{\text{lower}}(f_1)&:=[f_1-\hat{D}(f_1),f_1],
\label{eq:D_window}
\end{align}
 determine how much larger or smaller values $f_2$ can take given $f_1$. Therefore,

the useful purification window for a given fidelity, $f_1$, can be understood as the entire range of fidelities, $f_2$, that can yield a post purified fidelity greater ${\rm Max}(f_1,f_2)$ and is, therefore, given by,
\begin{align}
\Delta_{\rm use}(f_1)&=\Delta_{\rm upper}(f_1)\cup \Delta_{\rm lower}(f_1)\nonumber\\
&=[f_1-\hat{D}(f_1),f_1-\check{D}(f_1)].
\label{useful_window}
\end{align}

\begin{figure}
    \centering
    \includegraphics[width=\linewidth]{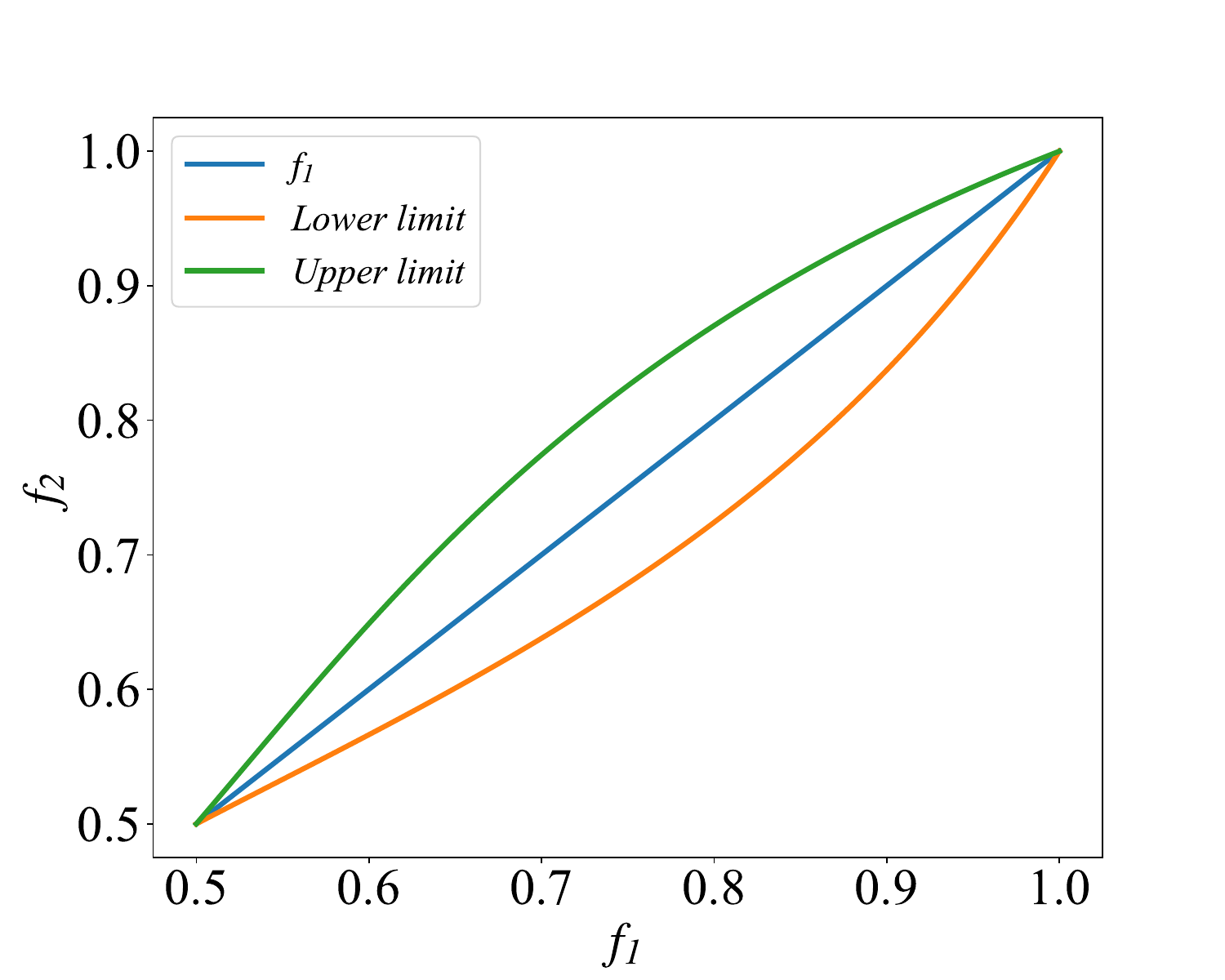}
    \caption{Fidelity window for useful entanglement purification. For a given $f_1$, $f_2$ could take any value in the region bounded by orange and green curves.}
    \label{window}
\end{figure}

While $f_1,f_2$ are random variables the quantities, $\check{D}(f_1),\hat{D}(f_1)$, are fully determined given the value of $f_1$ - as are the windows $\Delta_{\text{upper}}(f_1),\Delta_{\text{lower}} (f_1)$, and $\Delta_{\rm use}(f_1)$. The upper and lower limits of these useful windows are shown in Fig. \ref{window}.

In the situation that, $f_1\geq f_2$, Ineq. (\ref{ineq:D_up_down}) implies that the useful range of $f_2$ is, $f_2\in \Delta_{\text{lower}} (f_1)$; whereas, in the situation, $f_1\leq f_2$, the useful range is, $f_2\in \Delta_{\text{upper}} (f_1)$.

Several important qualitative instructions can be drawn from Ineq. (\ref{ineq:D_up_down}) and Eq. (\ref{eq:D_limits}) regarding the utility of MP-EP on MAD network paths between a given S-D pair.

First, note that because $|\check{D}(f_1)|\neq |\hat{D}(f_1)|$ in general, the fidelity windows $\Delta_{\text{upper}}(f_1),\Delta_{\text{lower}}(f_1)$ about a random path-fidelity, $f_1$, given by Eq. (\ref{eq:D_window}) are of unequal size on either side of $f_1$. This implies that the range of fidelities $f_2$ useful for purifying $f_1$ from below ($f_1\geq f_2$) is different from the range of fidelities $f_2$ that $f_1$ can be used to purify in the converse case ($f_2\geq f_1$).

Second, note that $\hat{D}(f_1)$ and $\check{D}(f_1)$ are non-monotonic functions of $F_1$ with both $\check{D}(f_1),\hat{D}(f_1)\to 0$ as $f_1\to 0.5,1$. The useful fidelity-window for purification $\Delta_{\rm use}(f_1)$ is therefore larger for intermediate ranges of the fidelity $0.5< f_1< 1$ than for low, $f_1\to 0.5$, or high, $f_1\to 1$ values of fidelity. An important implication is that fidelity-wise dissimlar states are useful for purification in the interior of the allowed fidelity range as compared to the extremes.

Finally, note that both $\check{D}(f_1)$ and $\hat{D}(f_1)$ approach zero as $f_1\to 1$. This is physically reasonable as full-rank states of the form (\ref{eq:network_state}) with high-fidelity can only be purified using another state with fidelity that is also high. This is markedly different from the situation when rank-2 states of the form $\rho=a\ket{\phi^+}\bra{\phi^+}+(1-a)\ket{\psi}\bra{\psi}, \psi=\phi^-,\psi^+,\psi^-$ are purified in which case the Ineq. (\ref{ineq:purif_cond}) is satisfied whenever $f_1,f_2\geq 0.5$ \cite{deutsch1996quantum,leone2021qunet}.

\subsection{Criteria for useful MP-EP along MAD paths}
\label{subsec:criteria}

Equipped with the expressions for the path parameter values, their probability distribution functions, and the fidelity window for useful purification we devise a set of two simple criteria to determine situations in which multipath entanglement purification can outperform the basic entanglement distribution protocol - on average. These two criteria,
\begin{align}\label{criteria}
\langle f_{\rm out}(f^{(l_0)},f^{(l_0+d)})\rangle\geq \overline{f^{(l_0)}},\tag{C1}\\
|\langle t^{(l_0)}\rangle-\langle t^{(l_0+d)}\rangle|\leq \tau_m,\tag{C2}
\label{criteria2}
\end{align}
respectively, are a condition on the post-purified fidelity (\ref{criteria}) and on the availability of network paths (\ref{criteria2}). We explain the physical implications of each of these criteria in the following paragraphs.

{\it Fidelity criterion:-} The fidelity criterion (\ref{criteria}) states the requirement that the expected post-purification fidelity of two paths of lengths, $l_0$ and $(l_0+d)$, be greater than the average fidelity of the shortest graph path (SGP) with length $l_{0}$ between the nodes S,D. The average on the l.h.s. is to be understood as the statistical average over the p.d.f.s of the fidelities of the paths involved in the MP-EP based protocol. When (\ref{criteria}) is satisfied for two MAD paths of length, $l_0$ and $(l_0+d)$, then MP-EP using MAD paths over those paths is advantageous compared to the basic entanglement distribution protocol over the path of length $l_0$. On the other hand, if this criteria is not satisfied by the two MAD paths, then MP-EP on those paths does not outperform the basic entanglement distribution protocol along the SGP between the two nodes. 

The utility of MP-EP as an entanglement distribution protocol for various distances between S,D nodes is revealed by criterion (\ref{criteria}). Expanding the l.h.s. of the inequality in terms of the p.d.f.s of the path fidelities, Eq. (\ref{eq:fidelity_pdf}), and the post purification output fidelity, Eq. (\ref{eq:fout}), as,
\small\begin{align}
&\int\limits_{0.5}^{1}\mathrm{d}f^{(l_0)}\int\limits_{0.5}^{1}\mathrm{d}f^{(l_0+d)}q_F^{(l_0)}(f^{(l_0)})q_F^{(l_0+d)}(f^{(l_0+d)})f_{\rm out}(f^{(l_0)},f^{(l_0+d)})\nonumber\\
&~~~~~~~~~~~~~~~~~~~~~~~~~~~~~~~~~~~~~~~~~~~~~~~~~~\geq \overline{f^{(l_0)}},\label{full_criteria1}
\end{align}\normalsize
we see that the inequality can be satisfied if the overlap of the path fidelity distibutions of the two paths, $q_F^{(l_0)}(f^{(l_0)})$ and $q_F^{(l_0+d)}(f^{(l_0+d)})$, in the useful purification window of $f^{(l_0)}$ given by Eq. (\ref{useful_window}) is sufficiently high - a situation that arises for suitable ranges of $l_0$ and $(l_0+d)$.

The implications of Ineq. (\ref{full_criteria1}) for the lengths of the paths, $l_0$ and $(l_0+d)$, are revealed by considering networks with the uniform edge fidelity distribution, $q _F(f )\sim U(f _{\rm min},1)$. In this case, the l.h.s. of Ineq. (\ref{full_criteria1}) scales as, $e^{-b(2l_0+d)}$, whereas, the r.h.s. can be approximated as $\simeq (1/4)+(3/4)e^{-cl_0}$, where, $b,c\geq0$ with the values of $b,c$ dependent on the value of $f _{\rm min}$ \footnote{This scaling holds for $1\leq l_0\leq 10$ which is the range of path lengths sufficient to cover even large random networks whose radius scales as $\ln |V|$.}. For a given, $l_0$, the l.h.s. of Ineq. (\ref{full_criteria1}) decreases exponentially with the difference of path lengths, $d$, and thus the inequality can only be satisfied for $d$-values upper bounded by a function of $l_0$ and $f _{\rm min}$. Conversely, for a fixed, $d$, the l.h.s. of the inequality also decreases exponentially with $l_0$ and can be satisfied only for $l_0$ up to a maximum determined by the values of $d$ and $f _{\rm min}$. 

\begin{figure}
    \centering
    \includegraphics[width=\linewidth]{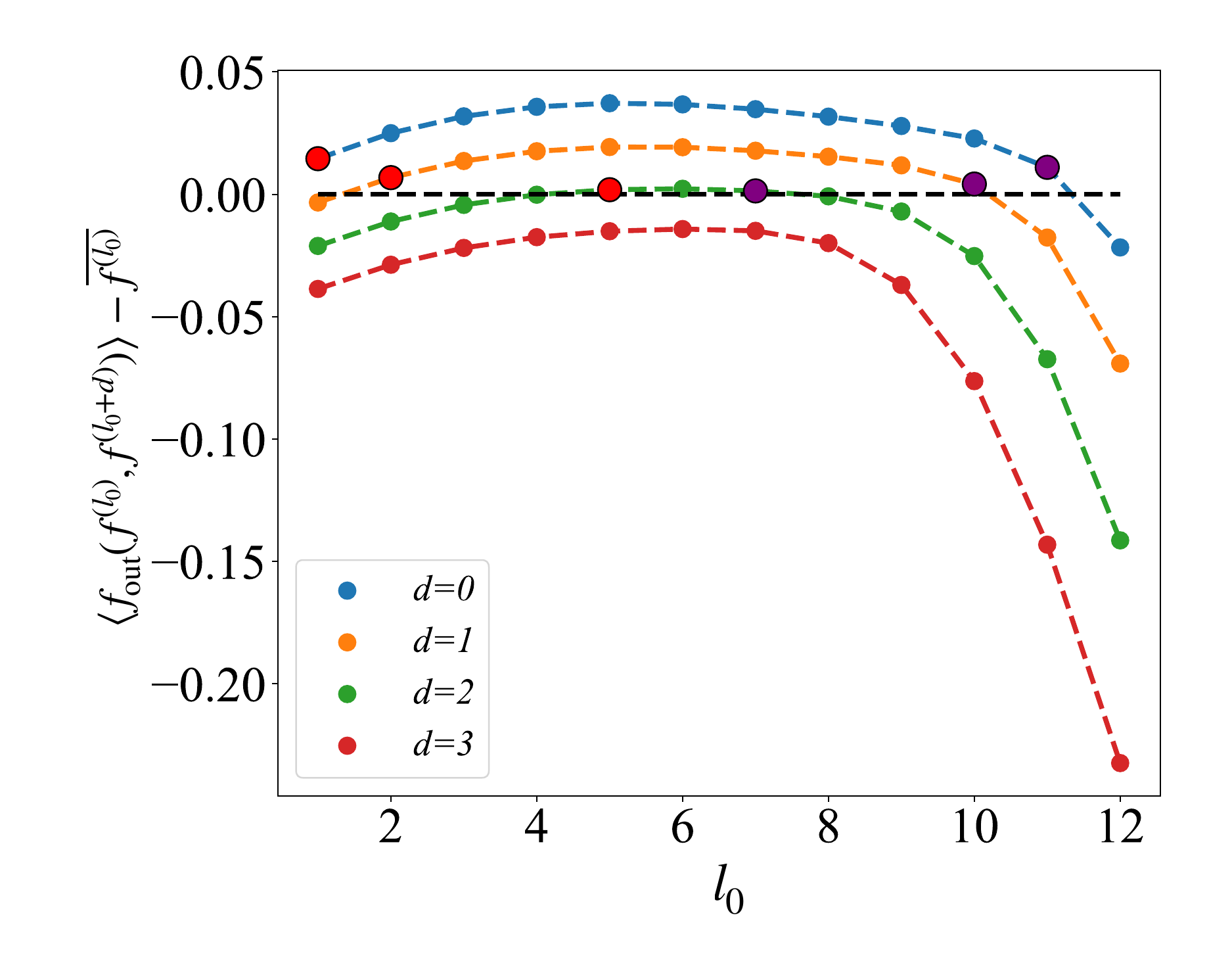}
    \caption{Difference of average fidelities obtained used the MP-EP protocol and the basic protocol for various shortest path distances, $l_0$, and the difference, $d$, of path lengths between S,D nodes in a network. The paths simulated for this plot had edge fidelities uniformly distribution between $[0.9,1]$. For a fixed $d$ the range of $l_0$ values for which the difference is positive indicates S,D separations where the expected fidelity using the MP-EP strategy performs better. The range of such $l_0$ values, that is, $l_0\in[l_*,l_{**}]$, is determined by the distribution of the edge parameter values. The Red and Purple highlighted points for every, $d$, denote the points corresponding to $l_*$ and $l_{**}$, respectively. The black dashed line represents the value of $l_0,d$, where, $\langle f_{\rm out}(f^{(l_0)},f^{(l_0+d)})\rangle= \overline{f^{(l_0)}}$.}
    \label{fig:useful_ld}
\end{figure}

For large separation between S,D nodes, $l_0\gg l_{\rm avg}$, compared to the average length scale of entangled paths $l_{\rm avg}$ in the network given by Eq. (\ref{eq:l_avg}), the r.h.s. of the Ineq. (\ref{full_criteria1}) converges to the value of, $1/4$, whereas, the l.h.s. of the inequality becomes exponentially small with $l_0$, thereby, suggesting a maximum separation $l_0$ between S,D where the inequality may be satisfied. Interestingly, for very small values of separation, $l_0=1,2,...\ll l_{\rm avg}$, between S,D the values of $b,c$ determined by $f _{\rm min}$ are such that the inequality (\ref{full_criteria1}) can only be satisfied when $d=0,1,...\ll l_{\rm avg}$ is also small. It turns out that the inequality (\ref{full_criteria1}) is satisfied for intermediate values of, $l_0\lesssim l_{\rm avg}$, for small values of $d$ as can be seen from Fig. \ref{fig:useful_ld}.

{\it Availability criterion:-} The second criterion (\ref{criteria2}) states the requirement that {\it both} the paths used for MP-EP need to be available within the coherence time of the quantum memories used in the network. The end-to-end entanglement distribution over the two paths succeeds probabilistically and the expected times of availability for the two paths are given respectively by, $\langle t^{(l_0)}\rangle=\langle1/p^{(l_0)}\rangle$ and $\langle t^{(l_0+d)}\rangle=\langle1/p^{(l_0+d)}\rangle$, where the average is taken with respect to the p.d.f. of the path probabilities, $q_P^{(l_0)}(p^{(l_0)})$ and $q_P^{(l_0+d)}(p^{(l_0+d)})$. Therefore, the availability requirement states that,
\begin{widetext}
\begin{align}\displaystyle
&\Bigg\vert{~\int\limits_{(p_{\rm min} )^{l_0}}^{1}\mathrm{d}p^{(l_0)}q_P^{(l_0)}(p^{(l_0)})\frac{1}{p^{(l_0)}} - \int\limits_{(p_{\rm min} )^{l_0+d}}^{1}\mathrm{d}p^{(l_0+d)}q_P^{(l_0+d)}(p^{(l_0+d)})\frac{1}{p^{(l_0+d)}} }\Bigg\vert\leq \tau_m.\label{new_criteria2}\tag{C2}
\end{align}
\end{widetext}
where, $\tau_m$ is the coherence time of the quantum memories used in the network. A reasonable scale for $\tau_m$ is given by $\tau_m\simeq (1/p_{\rm min})$ which correponds to the physical situation where the expected time for entanglement generation over the slowest edges in the network, $(1/p_{\rm min})$, are within the coherence time of the memories used.  

When the Ineq. (\ref{criteria2}) is satisfied by two paths of length, $l_0$ and $(l_0+d)$, the entangled states distributed over both the paths using the basic entanglement distribution protocol are available, on average, for MP-EP within the maximum waiting time allowed by the coherence time of the quantum memories. When this inequality is not satisfied, the two paths have widely different times of availability and are effectively unavailable together within the time $\tau_m$, thereby, prohibiting the use of MP-EP over MAD paths as a feasible entanglement distribution protocol.

The implication of Ineq. (\ref{criteria2}) for the lengths of the two paths, $l_0$ and $(l_0+d)$, can be seen by considering an approximation of the path probabilities as, $p^{(l_0)}\approx (\overline{p})^{(l_0)}$ and $p^{(l_0+d)}\approx (\overline{p})^{(l_0+d)}$. In this case, the l.h.s. of Ineq. (\ref{criteria2}) scales as, $e^{\log(1/\overline{p})l_0}(e^{\log(1/\overline{p})d}-1)$. Thus, for a fixed value of $d$, the l.h.s. increases with larger values of $l_0$ and the inequality cannot be satisfied for sufficiently large $l_0$. Conversely, for a given $l_0$ the l.h.s. increases with increasing values of $d$ and likewise implies a maximum value of $d$ up to which the inequality can be satisifed. 

\begin{figure}
    \centering
    \includegraphics[width=\linewidth]{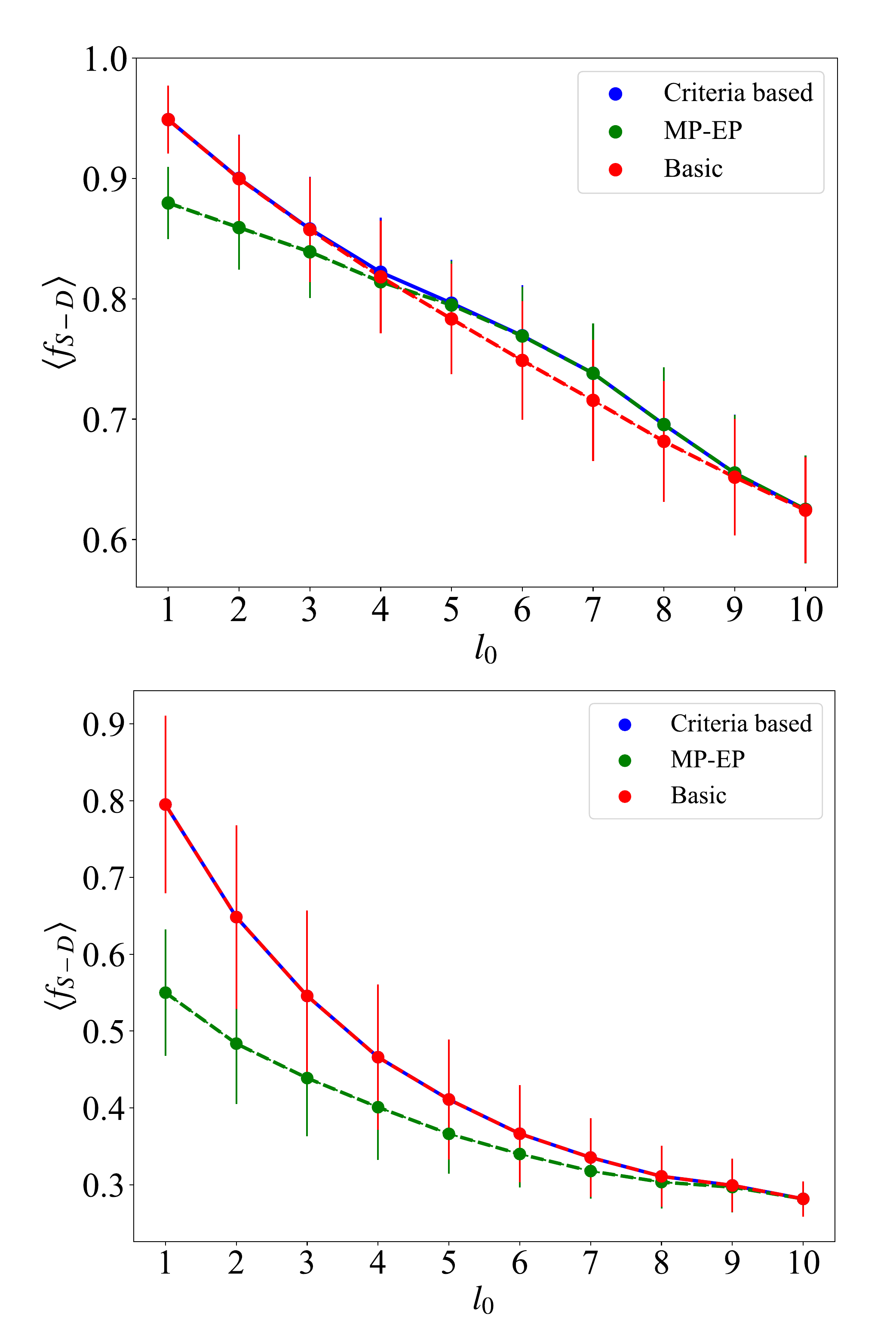}
    \caption{The average end-to-end fidelity, $\langle f_{S-D}\rangle$, between S-D node pairs vs the SGP length $l_0$ between the nodes for the basic, MP-EP and criteria based entanglement distribution protocols in a random network. The considered network contains $\abs{V}=10^4$ nodes and $\abs{E}=2.5\times10^4$ edges. In the plots, the Red and Green data points represent, $\langle f_{S-D}\rangle$, in the case of basic and MP-EP based protocols, respectively. The Blue data points indicate the average fidelity obtained using the criteria based protocol. In the top and the bottom panels the edge fidelity distributions are $U(0.9,1)$ and $U(0.6,1)$,  respectively. The edge probability distribution for both panels is $U(0.7,1)$. Note that the Blue data points overlap with the Red data points for $l_0\in\{1,2,3,4,9,10\}$ and with the Green data points for $l_0\in\{5,6,7,8\}$ in the top panel, whereas, they overlap with the Red data points for all $l_0$ in the bottom panel. The vertical bars represent the sample standard deviation of the end-to-end fidelity, $f_{S-D}$, for each of the protocols. Notice that the protocol based on the criteria (\ref{criteria}),(\ref{criteria2}) always chooses the better distribution strategy for all network scenarios and S,D separations.}
    \label{fig:mep_simulation}
\end{figure}

Together, the two criteria in (\ref{criteria}) and  (\ref{criteria2}) suggest that MP-EP using MAD paths of lengths $l_0,(l_0+d)$ between a pair of S,D nodes can be useful within a range of values, $l_0\in [l_*,l_{**}]$, for a given value of $d$. For random quantum networks the lengths of MAD paths between a pair of S,D nodes is determined by the graph topology, $G(V,E)$, and the value of $d$ can be different for distinct S,D pairs for the same value of $l_0$. The values of the minimum and maximum graph distance, $l_*$ and $l_{**}$ respectively, over which MP-EP outperforms the basic entanglement distribution protocol therefore depends on the specific network graph topology and can be estimated using numerical simulations of MP-EP given the network graph $G(V,E)$ and the edge parameter distributions $q _F(f )$ and $q _P(p )$.

\subsection{Numerical simulation of quantum network scenarios and validation of MP-EP criteria}
\label{subsec:numerical}
Numerical simulations of the basic entanglement distribution protocol and MP-EP using MAD paths on a random quantum network reveal the effectiveness of the criteria (\ref{criteria}) in determining the network scenarios as well as the appropriate source-destination distances where MP-EP can be usefully deployed as an entanglement distribution protocol, as shown in the two panels of Fig. \ref{fig:mep_simulation}. In both panels, the simulated network consisted of $|V|=10^4$ nodes and $|E|=2.5\times 10^4$ edges in a random topology. Each edge of the network was assigned a pair of parameters: the edge fidelity, $f$, and the edge probability, $p$, randomly chosen, respectively, from the edge-fidelity and edge-probability distributions, $q_F(f)$ and $q_P(p)$. 

Two network scenarios were considered. In the first the edge-fidelity distribution was chosen to have high average edge fidelity, $\overline{f}=0.95,~q _{F}(f )\sim U(0.9,1)$, whereas, in the second network scenario the average edge fidelity was lower, $\overline{f}=0.8,q_{F}(f)\sim U(0.6,1)$. In both scenarios the edge-probability distribution was kept fixed with, $q_{P}(p)\sim U(0.7,1),~\overline{p}=0.85$. For each scenario, a total of around $n_{\rm s}=10^4$ randomly chosen pairs of S,D nodes were sampled and for each pair, these four quantities were recorded:
\begin{enumerate}
\item Length of the shortest graph path between the S,D nodes.
\item Fidelity of shortest graph path between S,D according to the network topology, $G(V,E)$, using the basic entanglement distribution protocol.
\item Fidelity using MP-EP along two MAD paths, one of which was the shortest graph path with length $l_0$ and the other with length $(l_0+d)$, where, $d$ was the smallest allowed by the network topology $G(V,E)$. This post-purified fidelity was calculated irrespective of whether the two path lengths satisfied the criteria (\ref{criteria}). 
\item The satisfaction of the statistical criteria (\ref{criteria}) were checked using the values of the paths of length, $l_0$ and $(l_0+d)$. The value of the S,D-specific variable, ${\rm SAT(S,D)}$ was set to 1 if the criteria were satisfied otherwise it was set to 0 and recorded. Note that only the path lengths were used to evaluate the criteria but not the actual random path fidelities.
\end{enumerate}
The data from points 1 and 2 above was used to obtain the average fidelity over S,D pairs that had the same distance, $l_0$, among the sampled node pairs. The average fidelity, $\overline{f^{(l_0)}}$, for different values of $l_0=\{1,2,...,10\}$ was then plotted to obtain the scaling of average path fidelity vs the separation between S,D nodes achievable using the basic entanglement distribution protocol. These data points along with their empirical standard deviations are plotted in Red in the two panels of Fig. \ref{fig:mep_simulation} which show that the average fidelity decreases linearly in the high average edge fidelity scenario ($\overline{f}=0.95$), whereas, the decrease is much faster for low average edge fidelity, ($\overline{f}=0.8$).

With similar post-processing as above, the data from points 1 and 3 was used to obtain the scaling of the average fidelity when using MP-EP along two MAD paths vs the separation between S,D nodes which is shown in Green in the two panels of Fig. \ref{fig:mep_simulation}. This average fidelity is achievable when MP-EP is utilised along two available MAD paths between a S,D pair regardless of the MP-EP criteria (\ref{criteria}). Interestingly, from the top panel of the figure we see that MP-EP outperforms the basic protocol over a range of S,D separations, $5\leq l_0\leq 8$, but is not effective for larger separations, $l_0\to l_{\rm avg}$. At very short separations, $l_0\ll l_{\rm avg}$ the basic protocol, in fact, performs better than MP-EP for this network scenario. In the lower panel of the same figure with low average edge fidelities the basic protocol performs better than MP-EP for all values of the S,D separation, $l_0$.

The data from points 1,2,3 and 4 was used as follows. For all S,D pairs with the same value of $l_0$ the average fidelity was calculated by taking post-purified fidelities when ${\rm SAT(S,D)}=1$ and shortest graph path fidelities when ${\rm SAT(S,D)}=0$. The average fidelity in this case represents the expected fidelity between a pair of S,D nodes when the statistical criteria (\ref{criteria} )are utilised to choose between the basic- and MP-EP based entanglement distribution protocols. These data points are plotted in Blue in the two panels of Fig. \ref{fig:mep_simulation}. Noticeable from these two plots is that the criteria-based deployment of MP-EP always performs as well as the better of the basic and MP-EP protocols for a given separation of S,D nodes.

{\it Protocol performance in different network scenarios:-} The network scenario with high average edge fidelity, $\overline{f}=0.95, q _{F}(f )\sim U(0.9,1)$, is shown in the top panel of Fig. \ref{fig:mep_simulation}. For this network the average length of entangled paths was obtained using Eq. (\ref{eq:l_avg}) to be, $l_{\rm avg}=15$. The numerical data shows that MP-EP over two MAD paths, on average, outperforms the basic protocol by yielding a higher effective fidelity between S,D pairs separated by $l_0=5,6,7,8$. For this simulation, while the pair of MAD paths includes in all cases the shortest graph path with length $l_0$ between the pair of S,D nodes, the alternative path has length $(l_0+d)$ where $d$ is the smallest allowed by the network topology $G(V,E)$ for the particular S,D pair. In this particular network scenario, $l_*=5$, is the smallest separation while, $l_{**}=8$, is the largest separation between S,D pairs over which MP-EP can outperform the basic protocol. However, in general, the values of $l_*$ and $l_{**}$ depend on the network topology and edge parameter distributions. 

The network scenario with the lower average edge fidelity, $\overline{f}=0.8,~ q _{F}(f )\sim U(0.6,1)$, is shown in the lower panel of Fig. \ref{fig:mep_simulation}. For this network the average length of entangled paths was obtained using Eq. (\ref{eq:l_avg}) to be, $l_{\rm avg}=3$. The average path fidelities using the basic protocol over the entire range of path lengths, $l_0\in\{1,2,...,10\}$, was found to be higher than the average fidelity using MP-EP for the same separations between S,D pairs. Thus, the numerical data shows that MP-EP over MAD paths does not outperform the basic protocol for any range of $l_0$ values for such low average fidelity of the network edges. 

In any case, the MP-EP criteria presented in terms of the requirements (\ref{criteria}) can help identify the lengths of pairs of MAD paths, $l_0,(l_0+d)$, that can make MP-EP an advantageous protocol for entanglement distribution given statistical information about the edge parameters in terms their distributions $q _{F}(f )$ and $q _{P}(p )$. Conversely, the criteria can be used to choose between statistically better choice between the basic or MP-EP protocols.

\begin{figure}
    \centering
    \includegraphics[width=\linewidth]{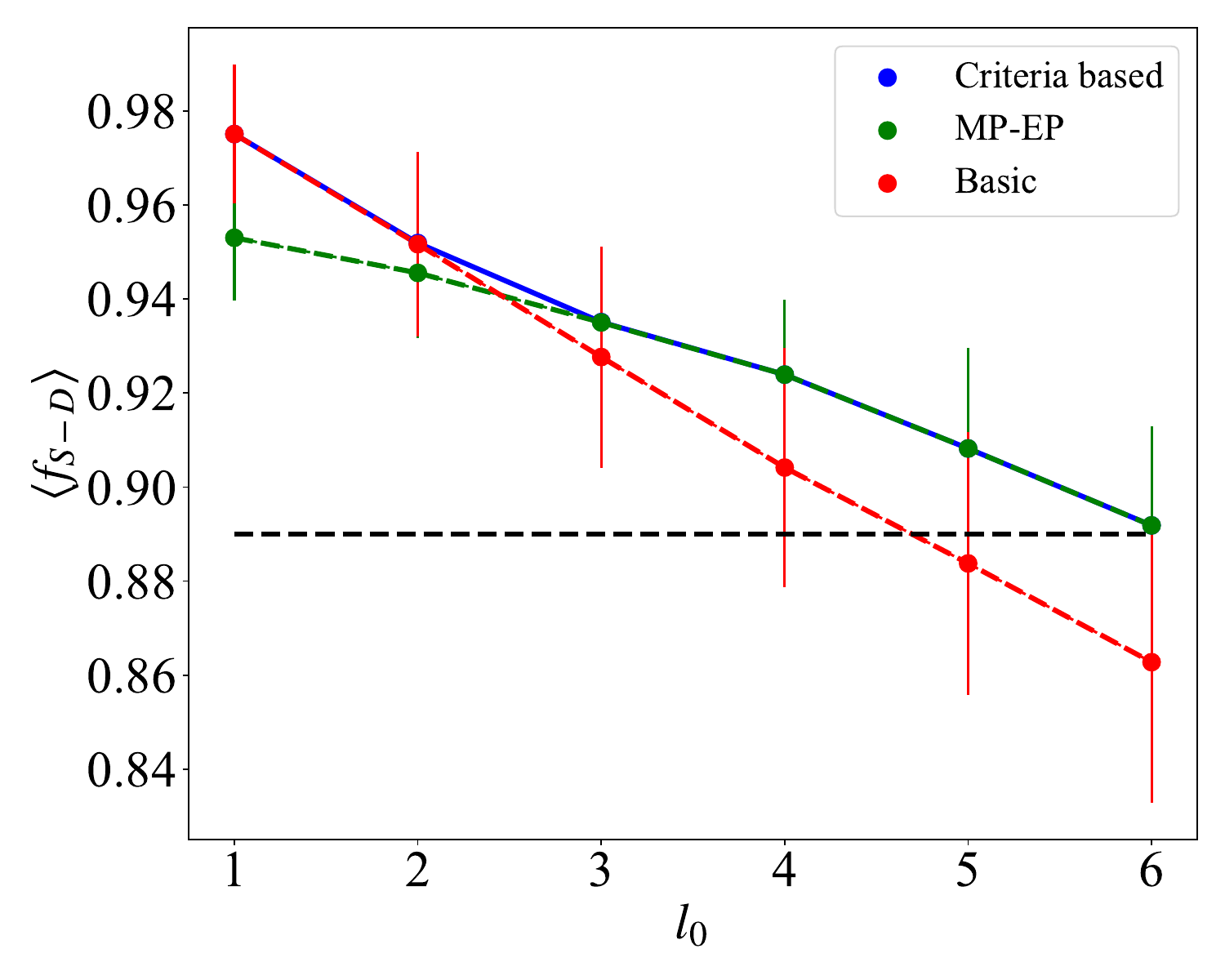}
    \caption{The average end-to-end fidelity, $\langle f_{S-D}\rangle$, plotted against the SGP length, $l_0$, simulated on a random network having $\abs{V}=10^4$ and $\abs{E}=5\times 10^4$ with an edge fidelity distribution $U(0.95,1)$ and edge probability distribution $U(0.7,1)$. The black dotted line represents the $\langle f_{S-D}\rangle=0.89$ line across different $l_0$. This represents the fidelity threshold, $f_{QKD}=0.89$, for QKD.}
    \label{fig:mep_boost}
\end{figure}

{\it Boosting effective fidelity of the entanglement connection for quantum tasks:-} When MP-EP does outperform the basic entanglement distribution protocol it can boost the effective fidelity between a pair of S,D nodes to reach the threshold required for a quantum task. We show an instance of this effect in Fig. \ref{fig:mep_boost} for a network with high average edge fidelity, $\overline{f}=0.975,~ q _F(f )\sim U(0.95,1)$, $|V|=10^4$ nodes and $|E|=5\times 10^4$ edges with a random topology, $G(V,E)$. From the figure, we find  that for, $l_0=5,6$, the average fidelity of paths using the basic protocol fails to reach the DI-QKD fidelity threshold of $f_{\rm QKD}=.89$ \cite{zhang2022device,zapatero2023advances}. On the other hand, the effective fidelity utilising MP-EP can be boosted to, $.91,.89$ for $l_0=5,6$ respectively, making DI-QKD feasible between those S,D nodes. In effect this brings a larger fraction of the network nodes within the region where DI-QKD can be performed increasing the functionality of the quantum network.

\begin{figure}
    \centering
    \includegraphics[width=\linewidth]{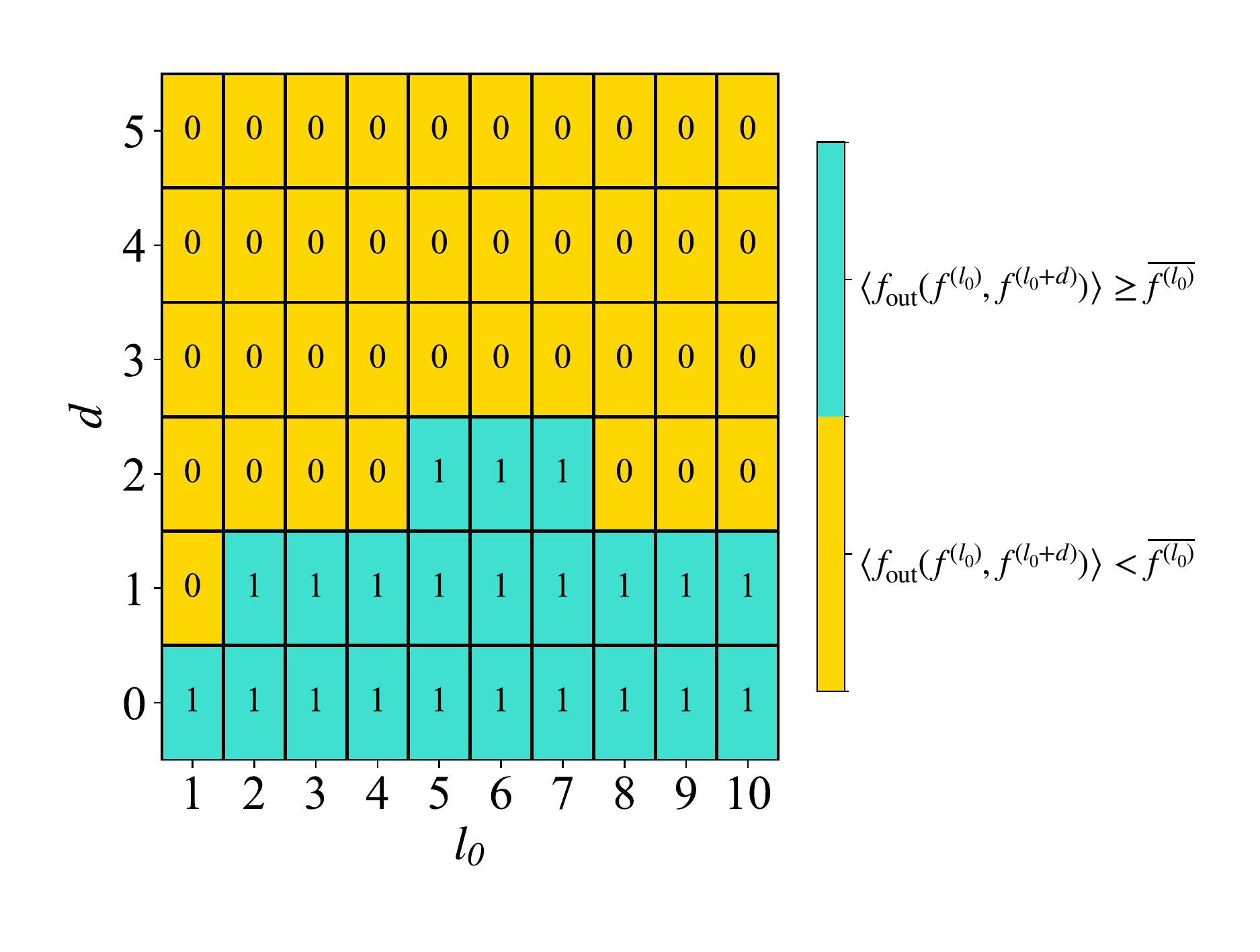}
    \caption{Decision table for performing MP-EP using two paths of lengths $l_0$ and $(l_0+d)$ based on the fidelity criterion, (\ref{criteria}). If the table element corresponding to a pair of $l_0$ and $d$ values is $1$ then MP-EP based protocol is advantageous compared to the basic protocol fidelity-wise. For all $0$ entries in the table the situation is reversed.}
    \label{fig:ld_heatmap_fidelity}
\end{figure}

\begin{figure}
    \centering
    \includegraphics[width=\linewidth]{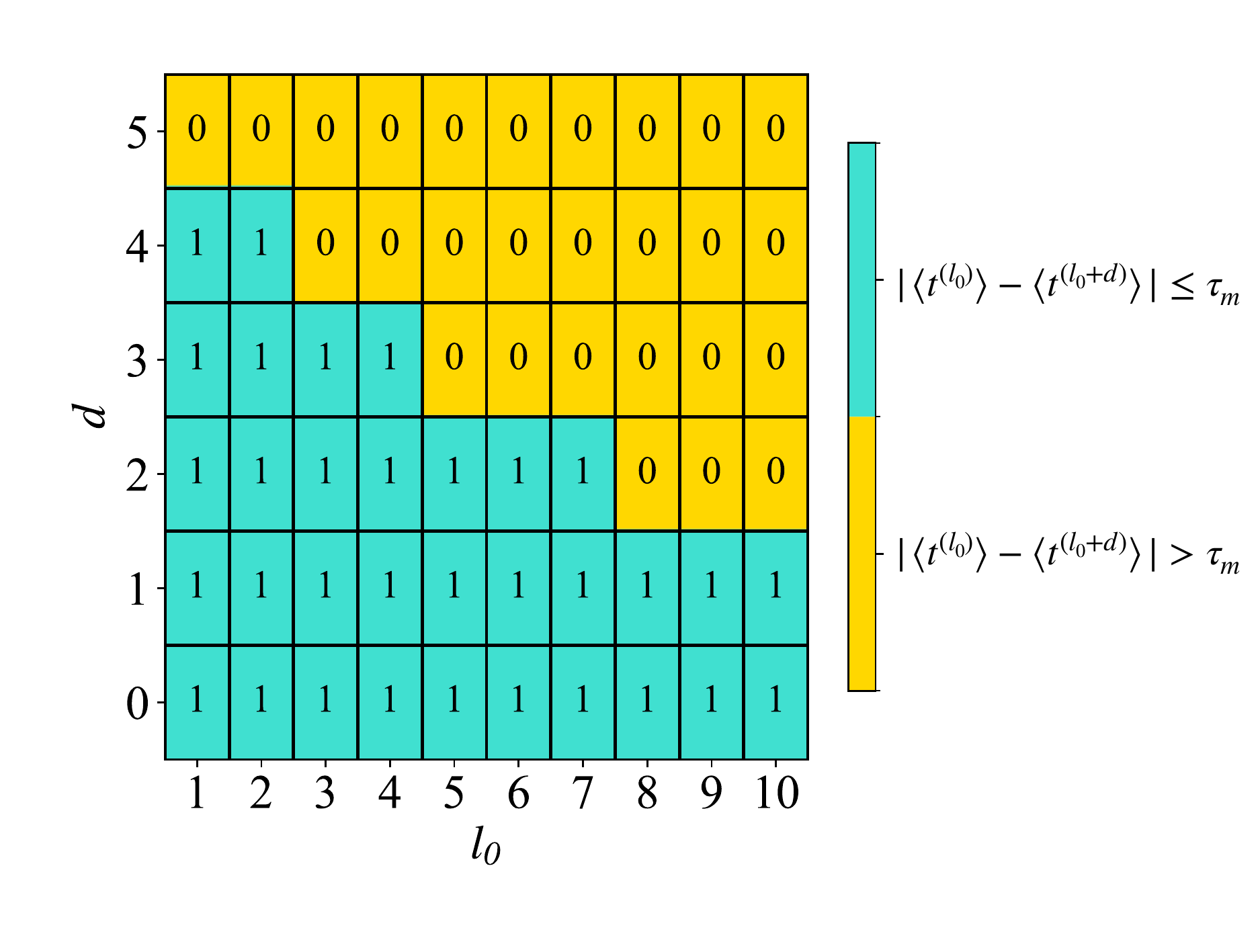}
    \caption{Decision table for performing MP-EP using two paths of lengths $l_0$ and $(l_0+d)$ based on the probability criterion, (\ref{criteria2}). If the table element corresponding to a pair of $l_0$ and $d$ values is $1$ then both paths are available within the lifetime of the quantum memories permitting the use of MP-EP. Whereas, for a $0$ entry the paths are not suitable for MP-EP.}
    \label{fig:ld_heatmap_prob}
\end{figure}

{\it Using MP-EP criteria for choosing between entanglement distribution protocols:-} Given the edge parameter distributions, 
$q _F(f )$ and $q _P(p )$, the evaluation of the fidelity and availability criteria (\ref{criteria}) for pairs of paths can be pre-computed and tabulated as shown in Figs. \ref{fig:ld_heatmap_fidelity} and \ref{fig:ld_heatmap_prob}. These tables can be used as a reference by a network controller to decide whether MP-EP should be used for a connection request between a pair of S,D nodes. If the network graph, $G(V,E)$, permits two MAD paths with lengths $l_0,(l_0+d)$ between the given S,D pair that corresponds to a pair of path lengths in the two tables which satisfy each of the two criteria in the respective tables then the controller can deploy MP-EP using these paths to obtain an effectively higher fidelity entanglement connection between the S,D node pair. In case the lengths of available paths, $l_0$ and $(l_0+d)$ fail to satisfy either of the criteria the controller can choose to use the entanglement only along the shortest graph path using the basic distribution protocol. This MP-EP criteria based choice of the entanglement distribution protocol as appropriate for a given S,D pair leads to a statistically optimal choice at all separations, $l_0$, between the nodes as shown via the Blue data points in Figs. \ref{fig:mep_simulation} and \ref{fig:mep_boost}.

\section{Discussion and conclusion}
\label{sec:discussion}
We presented an analysis of entanglement distribution based on a multipath entanglement purification strategy in Sec.  (\ref{sec:MAD}) using a statistical description of a quantum network, in Sec (\ref{sec:background}), which showed that MP-EP using MAD paths in a quantum network can be a useful protocol to strengthen the effective entanglement connection between a pair of source-destination nodes in a quantum network. We found that MP-EP using MAD paths can boost the entanglement fidelity above the thresholds required for a quantum task when the network edges have sufficiently high average fidelities and probabilities and the network graph topology allows MAD paths of appropriate lengths between suitably spaced node pairs. Further, we presented statistical criteria in Sec. (\ref{subsec:criteria}) to determine appropriate S,D separations in different network scenarios that can benefit from MP-EP and validated these criteria by simulating and comparing the single- and multpath- based entanglement distribution protocols on a quantum network with a random topology in Subsec. (\ref{subsec:numerical}).

We found that as the graph distance between S,D pairs, $l_0$, becomes large compared to the average length of entangled network paths, $l_{\rm avg}$, the fidelity of the entanglement connection using the basic protocol falls below the purifiable fidelity threshold of $0.5$. In this situation, $l_0\gg l_{\rm avg}$, the MP-EP strategy is of no benefit for increasing the fidelity of the entanglement connection. Conversely, when the spacing between the S,D pair is much smaller compared to the average length of entangled network paths, $l_0\ll l_{\rm avg}$, the MP-EP strategy generally does not perform as well as the basic protocol since the alternative paths have fidelities outside the useful purification window. Moreover, in this case single path entanglement distribution using the basic protocol is sufficient to obtain high fidelities, therefore, the MP-EP based protocol is not needed.

The entanglement distribution strategy of using MP-EP over MAD paths between S,D pairs in a quantum network is effective for node separations, $l_0\lesssim l_{\rm avg}$, comparable to the average length of entangled network paths under appropriate conditions of the network graph topology and parameter distributions. Broadly, quantum networks with high average edge fidelities and probabilities and topologies that allow MAD paths of similar lengths between S,D pairs are well suited to benefit from the MP-EP strategy as shown in Subsec. (\ref{subsec:numerical}). 

The technique of comparing the MP-EP and basic distribution protocols using a statistical model of the quantum network presented here can be extended to reveal the advantage of using multipath strategies for more sophisticated protocols for entanglement distribution using single network  paths such as linear quantum repeater protocols \cite{muralidharan2016optimal,RevModPhys.95.045006}. To do so, one would need the distribution functions of network path parameters that can be obtained by numerical means \cite{li_elkouss} for a given entanglement distribution protocol and the distribution functions of network edge parameters of the quantum network. 

As quantum networks evolve in size and complexity, entanglement distribution protocols that exploit its complex-network properties will be useful to increase their functionality. Towards this goal it would be important to investigate the advantage provided by MP-EP in quantum networks with different topologies such as regular or scale-free networks and studying the MP-EP strategy when the set of multiple paths used in the MP-EP strategy has a large number of paths \cite{sohel_mep}. Purifying the entanglement resources provided independently by this set of MAD paths, as permitted by the network topology, can potentially significantly strengthen the entanglement connection between arbitrary network nodes and enable implementation of quantum tasks over large-scale quantum networks.

{\it Acknowledgement:-} Funding from DST, Govt. of India through the SERB grant MTR/2022/000389, IITB TRUST Labs grant DO/2023-SBST002-007 and the IITB seed funding is gratefully acknowledged. We would like to thank David Elkouss, Bill Munro and Nicolo Piparo for useful conversations.
\newpage
\appendix


\section{Probability distribution of path fidelity}
\label{app:pdf_path_fidelity}
Entanglement connection between any two nodes S,D is realised by performing entanglement swapping at all intermediary nodes. This implies that if the two nodes are separated by a distance $l$, $(l-1)$ swappings are performed to share an entangled state of fidelity $f^{(l)}$ between them. The fidelity $f^{(l)}$ follows the relation  as given in the Eq. (\ref{eq:swapping_fidelity}), which we re-write here for the sake of completeness:
\begin{equation}
    f^{(l)}=\frac{1}{4}+\frac{3}{4}\prod_{i=1}^l\Big(\frac{4f_i -1}{3}\Big).
\end{equation}
Here, the symbol $f_i $ represents the fidelity of the shared state in the $i^{\rm th} $ pair or $i^{\rm th}$ edge. 

To determine the p.d.f. of path-fidelity, we re-express the path-fidelity using $\log$ function as,
\begin{equation}
    \log{\Big(\frac{4f^{(l)}-1}{3}\Big)}=\sum_{i=1}^l \log{\Big(\frac{4f _i-1}{3}\Big)}
\end{equation}
The above equation suggests that the p.d.f. of l.h.s. can be obtained as an $l$-fold convolution of p.d.f of ${\small \log{\Big(\frac{4f _i-1}{3}\Big)}}$ which can then be inverted to find p.d.f. of $f^{(l)}$.

    To do so, we follow the following steps in order.
\begin{itemize}
\item Determine  the p.d.f. $q _Y(y)$ of $Y=\log{\Big(\frac{4f -1}{3}\Big)}$ using $q_F (f )$.
\item Employ Convolution theorem  and Fourier transform techniques to find  the p.d.f. of $Z=\log{\Big(\frac{4f^{(l)}-1}{3}\Big)}$. 
\item Invert the p.d.f. $q_Z(z)$ to find the p.d.f. $q_F^{(l)}(f^{(l)})$.
\end{itemize}

To do so, we consider the edge fidelity to be uniformly distributed between  $f _{\rm min}$ and $1$ in which case its p.d.f. is given as:
\begin{equation}
    q_F (f )=\frac{1}{1-f _{\rm min}}.
\end{equation}
As a next step, we determine the p.d.f. of $Y$, $q_Y(y)$ using change of variable techniques and its Fourier transform ${\cal F}( q_Y)$ as follows:
\begin{align}
   & q_Y(y)=\frac{3}{4}q_F \exp{y},\quad\quad \log{\Big(\frac{4f _{\rm min}-1}{3}}\Big)\leq y\leq 0.\\
   &{\cal F}( q_Y(y))=\frac{3}{4}q_F \Bigg(\frac{1-\exp{(1+i\omega)\log{\big(\frac{4f _{\rm min}-1}{3}\big)}}}{1+i\omega}\Bigg)
\end{align}
We, next, employ Fourier convolution theorem \cite{arfken2011mathematical} which states that for a function of the form $Z=\sum_iY_i$, the fourier transform of $q_Z(z)$ is the product of fourier transforms of $q_{Y_i}(y_i)$. That is,
\begin{equation}
    {\cal F}(q_Z)=\prod_i{\cal F}(q_{Y_i}).
\end{equation}
Following this and taking inverse fourier transform, we identify $q_Z(z)$ to be,
\begin{align}
    q_Z(z)=&\frac{-1}{(l-1)!}\big(\frac{-3}{4}q_F \big)^l\exp{z}\nonumber\\
    &\sum_{n=0}^{m-1}\binom{l}{n}(-1)^n\big(n\log(\frac{3}{4f_{\rm min} -1})+z\big)^{l-1},
\end{align}
where $m\log{\big(\frac{4f_{\rm min} -1}{3}\big)}\leq z\leq (m-1)\log{\big(\frac{4f_{\rm min} -1}{3}\big)}$ and $m=\{1,2,\cdots, l\}$.
This allows to find the p.d.f.  of $q_F^{(l)}(f^{(l)})$ by again employing change of variable technique,
\begin{align}
    q_F^{(l)}(f^{(l)}=&\frac{1}{(l-1)!}\Big(\frac{3}{4}\Big)^{l-1}\Big(\frac{1}{(1-{f_{\rm min} }}\Big)^l\nonumber\\
    &\sum_{n=0}^{m-1}(-1)^n\binom{l}{n}\Bigg[\log{\bigg(\frac{(4f_{\rm min} -1)^n}{(4f^{(l)}-1)3^{n-1}}\bigg)}\Bigg]^{l-1}
\end{align}
where, the random path fidelity $f^{(l)}$ can lie in one of the `$l$' geometrically-spaced intervals of path fidelity, 
$$\frac{1}{4}+\frac{3}{4}\bigg(\frac{4f_{\rm min} -1}{3}\bigg)^m\leq f^{(l)}\leq \frac{1}{4}+\frac{3}{4}\bigg(\frac{4f_{\rm min} -1}{3}\bigg)^{m-1}$$ and $m\in\{1,2,\cdots,l\}$ as discussed in Sec. (\ref{subsec:random_path_parameters}).

\section{Identifying useful fidelity window}
\label{app:window}
Consider two states of the isotropic form shared between $S$ and $D$ nodes along two paths $R_1$ and $R_2$ be:
\begin{align}
    &\rho_1=\frac{(4f_1-1)}{3}\ket{\phi^+}\bra{\phi^+}+\frac{1-f_1}{3}\mathbb{1},\\
    &\rho_2=\frac{(4f_2-1)}{3}\ket{\phi^+}\bra{\phi^+}+\frac{1-f_2}{3}\mathbb{1}
\end{align}
If the two states are purified using the Deutsch's purification protocol, the output state will be diagonal in the Bell-basis with the output fidelity as:
{\small \begin{align}
     &f_{\rm out}(f_1,f_2)=\nonumber\\
     &\frac{f_1f_2+\frac{1}{9}(1-f_1)(1-f_2)}{f_1f_2+{\frac{1}{3}(f_1(1-f_2)+(1-f_1)f_2})+\frac{5}{9}(1-f_1)(1-f_2)}.
\end{align}}
Purification is termed successful only if the output fidelity is the maximum of the input fidelity, i.e.,
\begin{equation}\label{eq:use_puri}
    f_{\rm out}\geq{\rm max.}(f_1,f_2)
\end{equation}
The two cases arise: (i) $f_1\geq f_2$, and (ii)  $f_2\geq f_1$.
In the first case, the condition of successful purification, Eq. (\ref{eq:use_puri}) reduces to
\begin{align}
    f_{\rm out}\geq f_1
\end{align}
which upon simplification gives:

\begin{align}
     &f_2\geq \dfrac{2f_1^2-6f_1+1}{8f_1^2-12f_1+1}\nonumber\\
  &f_1-f_2 \leq \dfrac{8 f_1^3-14 f_1^2+7 f_1-1}{8f_1^2-12 f_1+1}
\end{align}
Since $D(f_1,f_2):=f_1-f_2$, it implies
\begin{align}
 &   D(f_1,f_2)\leq \dfrac{8 f_1^3-14 f_1^2+7 f_1-1}{8f_1^2-12 f_1+1}\nonumber\\
 & D(f_1,f_2)\leq \hat{D}(f_1)
 \label{eq:upper}
\end{align}
where $\hat{D}(f_1)=\dfrac{8 f_1^3-14 f_1^2+7 f_1-1}{8f_1^2-12 f_1+1}$.
In a similar manner, in the second case ($f_2\geq f_1$ ), the condition of useful purification reduces to 
 \begin{equation}
    f_{\rm out}\geq f_2,\nonumber
\end{equation}
which upon simplification provides,
\begin{align}
    f_2^2(8f_1-2)+f_2(6-12f_1)+f_1-1\leq 0
\end{align}
Re-expressing the above inequality, we get: 
\begin{align}\label{eq:f2}
    &(f_2-f_2^+)(f_2-f_2^-)\leq 0\implies f_2\leq f_2^+
\end{align}
 where $f_2^\pm=\dfrac{(6f_1-3)\pm\sqrt{28f_1^2-26f_1+7}}{2(4f_1-1)}$. 
 The  inequality (\ref{eq:f2}) takes the form:
 \begin{align}
      &f_2\leq  \dfrac{(6f_1-3)+\sqrt{28f_1^2-26f_1+7}}{2(4f_1-1)},
 \end{align}
 which can be re-written as the difference of path-fidelities as follows:
\begin{align}
   &D(f_1,f_2)= f_1-f_2\geq  f_1 - \dfrac{(6f_1-3)+\sqrt{28f_1^2-26f_1+7}}{2(4f_1-1)}\nonumber\\
  & D(f_1,f_2) \geq \dfrac{8f_1^2-8f_1+3-\sqrt{(5f_1-2)^2+3(1-f_1)^2}}{8f_1-2}\nonumber\\
   & D(f_1,f_2)\geq \check{D}(f_1)
   \label{eq:lower}
\end{align}
where $\check{D}(f_1)=  \dfrac{8f_1^2-8f_1+3-\sqrt{28f_1^2-26f_1+7}}{8f_1-2}$.
Clubbing the two inequalities (\ref{eq:upper}) and (\ref{eq:lower}), we got the region of useful purification as:
\begin{align}
    \check{D}(f_1)\leq D(f_1,f_2)\leq \hat{D}(f_1).
\end{align}
which is the same as given in inequality (\ref{ineq:D_up_down}).
\section{Simulation of MP-EP on MAD paths in random quantum networks}
\label{app:algorithm}
To simulate MP-EP on MAD paths in random quantum networks, random graphs $G(V,E)$ with the desired number of vertices $|G|$ and edges $|E|$ are generated. To generate the data for average path fidelity, pairs of randomly chosen S,D pairs are sampled. For a randomly chosen pair of S,D nodes the shortest graph path with length $l_0$ is obtained using Dijkstra's algorithm. To find an alternative and distinct path which is edge-disjoint from the previously found path a modified network graph $G'(V,E')$ with the edges contained in the previous path removed from the set of edges of the new graph is created. This can be iterated to find the entire set of MAD paths between a pair of S,D nodes up to a maximum length. Here, we only identify two MAD paths with the first being the shortest graph path between S,D.

The computational complexity of this strategy scales linearly with the size of the set of MAD paths, $\mathcal{R}$ defined in Eq. (\ref{MAD_Paths}), and the complexity of Dijkstra's algorithm. Specifically, for a network graph $G(V,E)$, the total complexity of finding $|\mathcal{R}|$ MAD paths is given by,
\begin{align}
    \mathcal{O}\left( |\mathcal{R}| \cdot \left( \lvert E \rvert + \lvert V \rvert \log \lvert V \rvert \right) \right),
    \label{complexity_path_finding}
\end{align}
which reflects the cost of running Dijkstra's algorithm $|\mathcal{R}|$ times, each time updating the graph by removing edges corresponding to the previously found path.

After identifying 2-MAD paths between a specific source-destination (S-D) node pair, the random fidelity of the end-to-end entangled state obtained along the two paths by swapping the random states over the edges is calculated. Using the fidelities of these two states, the post-purification fidelity is calculated. 



\begin{algorithm}[H]
\caption{Multi-Path Entanglement Purification (MP-EP) Algorithm}
\begin{algorithmic}[1]
\State \textbf{Input:} Network $G$.
\State \textbf{Output:} Multipath purified fidelity, $f_{\rm out}(f_1,f_2)$.
\vspace{0.2cm}

\State \textbf{Define:} Dijkstra(S, D)
\Comment{Obtain the SGP between two nodes.}

\State \textbf{Define:} MP-EP(path1, path2)
\Comment{Performs MP-EP using path1 and path2.}

\vspace{0.2cm}
\State \textbf{function Find\_Paths(S,D,k):}
\State \hspace{0.5cm} Initialize $Paths \gets []$
\For{$num\_paths \gets 1$ \textbf{to} $k$}
    \State $SGP \gets$ Dijkstra(S, D) 
    \State $Paths$.append($SGP$) 
    \State $G$.remove\_edges($SGP$.edges()) 
\hspace{0.5cm}
\EndFor
\State \hspace{0.5cm} \textbf{return} $Paths[0],\dots,Paths[k-1]$
\State \textbf{end function}
\vspace{0.2cm}

\For{$N_S \gets 1$ \textbf{to} 100}
    \State $S \gets$ random node from $G$
    \For{$N_D \gets 1$ \textbf{to} 100}
        \State $D \gets$ random node from $G$
        \State Path1, Path2 $\gets$ Find\_Paths(S, D, $2$) 
        \State $f(\mathcal{P}(f_1,f_2))=$ MP-EP(Path1, Path2)
    \EndFor
\EndFor
\end{algorithmic}
\end{algorithm}

%

\end{document}